\documentclass[aps,prc,preprint]{revtex4-1}
\usepackage{graphicx}
\usepackage{xcolor}
\usepackage{amsmath}
\usepackage{mathtools}
\usepackage{wasysym}
\usepackage{multirow}
\begin{document}
\title{Structural Properties of Rotating Hybrid Compact Stars with  Color-Flavour-Locked 
Quarks Matter core and its Tidal Deformability}
\author{Suman Thakur}
\email{sumanthakur88@gmail.com}
\author{Shashi K. Dhiman}
\email{shashi.dhiman@gmail.com}
\affiliation{Department of Physics, Himachal Pradesh University, Summer-Hill,
Shimla - 171005, India.}
\date{\today}
\begin{abstract}
We investigate the possible sequels, for the hybrid compact stars consisting of nucleons,
hyperons and three flavour color-flavor-locked quarks phase under global 
neutrality and chemical equilibrium conditions. The hadron equations of state
are computed within the framework of energy density functionals based on the 
relativistic mean field theory by employing two different model parameterizations. 
The quarks matter phase of equation of state is computed by using Quarks 
Quasiparticle model derived from a non-relativistic energy density-functional 
approach. A plausible set of hybrid equations of state for superdense hadron-quarks
matter is obtained and, which satisfies the constraints provided by bulk nuclear matter, 
the observational data of frequencies and maximum Gravitational mass 2$M_\odot$ 
with their extracted radii, Keplerian limits, secular axisymmetric instability
and limits of tidal deformabilities from GW170817 binary neutron stars (BNS) merger. 
We obtained structural properties of nonrotating and rotating compact stars, 
the evolutionary sequences with constant baryonic mass star spinning 
down by electromagnetic and gravitational radiations and, discuss the issues 
relating to phase transition from hadron to quarks matter phase. 
The internal structure of rotating star with observed spin down frequencies,
exhibiting shrinkage of soft quarks core of compact stars are discussed for 
constant baryonic mass. We present the theoretically computed limits of 
radii for the spin down configurations of hybrid stars 
corresponding to the recently observed millisecond pulsars.
We also present and discuss the calculated results for tidal deformability,
$\Lambda$ and tidal parameters $\Lambda_1$,  $\Lambda_2$ 
of non spining hybrid star which are consistent
with the waveform model analysis of GW170817 BNS.
\end{abstract}
\maketitle
\section{Introduction}
\label{intro} 
 It has been investigated \cite{Witten1984,Farhi1984,Haensel1986,Alcock1986} that 
 the dense matter in the core of compact stars may consist of quark matter and,
 should be described as a color quarks as the fundamental degree of freedom. 
 Therefore, the hybrid stars phenomenology offers a unique tool to address the 
 challenge of understanding the phase transition in the dense quantum chromodynamics 
 (QCD). Recently, the nuclear theory studies \cite{Haensel2007,Lattimer2014,Baym2018} 
are mainly in focus for understanding dense matter of compact stars (CS). 
Recent observations at LIGO and Virgo of GW170817 event \cite{Abbott2018,Abbott2019}
identified as almost certainly the merger of Binary Neutron Stars
and the discovery of CS with masses around 2$M_\odot$ \cite{Demorest2010,
Antoniadis2013,Arzoumanian2018} have intensified the interest in these 
intriguing objects. The analysis GW170817 has demonstrated the potential
of gravitational wave (GW) observations to yield new information relating to 
the limits on CS tidal deformability. In addition to these astrophysical 
observations \cite{Champion2008,Abdo2010,Guillemot2012,Fonseca2016,Ozel2016}
the measurements of rotation frequencies  of the pulsar can be employ to 
constrain the composition and behaviour of Equations of State (EOS) of the
dense nuclear matter. However, the direct measurement
of radius of CS is still a great challenge from astrophysical observations.
The upcoming high-precision x-ray space missions, such as the ongoing NICER 
(Neutron Star Interior Composition Explorer) \cite{Gendreau2012,Fonseca2016,
Arzoumanian2018} and the future eXTP (Enhanced X-ray Timing
and Polarimetry Mission) \cite{Zhang2014} have aimed to improve the situation by
simultaneous measurements of CS masses and radii with higher
accuracy \cite{Watts2016,Oertel2017}. It is also expected that limits on CS radii
are to be improved by new detections of gravitational wave signals from neutron 
star mergers.
\par The investigations of the observed masses and extraction of the radii of CS would
allow us to reveal composition and various phases of EOS of dense matter. Many attempts 
\cite{Dhiman2007,Ozel2016,Oertel2017} have been made by using microscopic many body nuclear
theories to model the nuclear matter EOS capable of constructing CS, containing nucleons,
hyperons and quarks, under the constraint of global $\beta$-equilibrium. 
The EOS including hyperons and/or quarks are typically much softer than those containing
just nucleons, leading to the reduction of maximum Gravitational masses of CS which may 
not be compatible with recent observations of 2$M_\odot$. Whereas, the properties of the
composition of EOS of CSs are not completely understood due to an apparent complications
of non-perturbative effects of QCD. In the low energy regime the QCD coupling constant 
will be significantly large and experimental data have large uncertainties 
\cite{Bethke2007}, therefore, it is not yet possible to obtained a reliable 
EOS of quark matter phase from the first principle of QCD. The quarks matter phase of 
EOSs have been treated by employing phenomenological
models with some basic features of QCD, such as, 
the MIT bag models \cite{Alford2005,Agrawal2009,Zhou2018} with a bag constant and
appropriate perturbative QCD corrections and 
Nambu-Jona-Lasino with chiral symmetry and its breaking \cite{Menezes2006}.  
\par In the present work we compute a set of hybrid equations of state, where the hadron
phase has been calculated within the framework of energy density functionals based
on the relativistic mean field theory \cite{Dhiman2007} and, the quarks matter phase
of equation of state is computed by using Quarks Quasiparticle model QQPM \cite{Xia2014} 
with CFL phase \cite{Agrawal2009}. The medium effects are included in the cold 
quarks matter in terms of variation in effective mass of quarks 
and effective bag parameters as function of chemical potential.  
Further a plausible set of hybrid equations of state for hadron-quarks
matter is employed to determine the maximum Gravitational mass, equatorial
radius and rotation frequency of stable stellar configuration of hybrid stars, 
which satisfies the constraints provided by Keplerian limits, secular axisymmetric
instability, the observational data of frequencies and maximum Gravitational mass 
2$M_\odot$ of static sequences with their radii and GW170817 event of neutron stars 
merger. We further constrain the radii of hybrid star to the recent limit extracted 
from GW170817 BNS  
R$_{1}$=11.9$^{+1.4}_{-1.4}$km and R$_{2}$ =11.9$^{+1.4}_{-1.4}$km.
We presented the evolutionary sequences for hybrid CS with constant baryonic mass spinning
down by electromagnetic and gravitational radiations and, discuss the issues 
relating to phase transition from hadron to quarks matter phase. 
The internal structure of rotating star with observed spin down frequencies, 
exhibiting shrinkage or expansion of soft core of star are discussed for constant
baryonic mass. We present the theoretically computed limits for the results of spin
down configurations of hybrid stars corresponding to the recently observed millisecond 
pulsars, for the Love number, k$_{2}$ and the tidal deformability, $\Lambda$.
We approximate the hybrid star as a axisymmetric and rigid rotating body, and 
resort to Einstein's theory of general relativity for a rapidly rotating star. 
Recently, numerical methods for (axisymmetric) rotating stellar 
structure have been advanced by several groups \cite{Komatsu1989, Weber1991, 
Cook1992, Cook1994, Salgado1994, Stergioulas1995, Stergioulas2003}. 
In this work we employed the Rotating Neutron Star RNS method and code 
\cite{Komatsu1989,Stergioulas1995} to calculated the properties of rapidly
rotating Hybrid CSs.
\par The paper has been organised as, in section II, we described the theoretical
framework employed to compute the hybrid EOSs. The Extended field-theoretical 
relativistic mean field (EFTRMF) models have been employed to 
describe the nucleons and hyperons phases and, whereas the quarks matter
phase has been obtained from the QQPM.
The mixed phase of hybrid equations of state is obtained by using Glendenning 
construction based on Gibbs conditions of equilibrium. The framework of Relativistic Rotation of CSs are
described in section III. The section IV, present results
and discussions for equations of state employed to construct the non-rotating
and rapidly rotating hybrid stars.
The rotational evolution for a constant baryonic mass is also discussed.
We have also presented and discussed the theoretical results of the tidal
deformability of hybrid star in the context of GW17017 event of Neutron Stars merger. 
The conclusions of the present research works are presented in section V.         
\section{Theoretical Framework}
In this section, we discuss the theoretical approaches employed to the calculated
sets of equations of state of dense nuclear matter in different phases.
The EFTRMF model parameter BSR3 has been successfully applied in describing the 
nuclear structure properties of finite nuclei, properties of bulk nuclear matter 
at saturation densities, dense asymmetric nuclear matter, hadrons and hyperons nuclear
matter at high densities \cite{Dhiman2007}. These model parameters have been adopted 
to compute EOSs and construct neutron stars and hybrid CSs.
The final hybrid EOS is comprised of two separate EOSs for 
each phase of matter, which are combined by utilising a Glendenning phase transition
construction \cite{Glendenning1992,Glendenning2000}.
\subsection{Hadron Phase}
 In the EFTRMF model the effective Lagrangian density consists of
self and mixed interaction terms for $\sigma$, $\omega$, and $\rho$ mesons upto
the quartic order in addition to the exchange interaction of baryons with $\sigma$, 
$\omega$, and $\rho$ mesons. The $\sigma$, the $\omega$, and the $\rho$ mesons are 
responsible for the ground state properties of the finite and heavy nuclei. 
The mixed interactions terms containing the $\rho$-meson field enables us to vary
the density dependence of the symmetry coefficient and neutron skin thickness in
heavy nuclei over the wide range without affecting the other properties of the 
finite nuclei \cite{Furnstahl2002,Sil2005}. In particular, the contribution from 
the self-interaction of $\omega$-mesons peforms an important role in determining the
high density behaviour of EOS and structure properties of compact stars 
\cite{Dhiman2007,Muller1996}. Whereas the inclusion of self-interaction of 
$\rho$-meson affect the ground state properties of
heavy nuclei and compact stars only very marginally \cite{Muller1996}. 
 In the present work, We use only BSR3 parameterization to construct the hadronic
 phase of the Hybrid EOS. 
 The Lagrangian density
for the EFTRMF model can be written as
\begin{equation}
\label{eq:lden1}
{\cal L}= {\cal L_{BM}}+{\cal L_{\sigma}} + {\cal L_{\omega}}
+ {\cal L_{\mathbf{\rho}}} + {\cal L_{\sigma\omega\mathbf{\rho}}} +
{\cal L}_{em}+ {\cal L}_{e\mu} +  L_{YY},
\end{equation}
A detailed description of each term of the effective Lagrangian density
Eq.(\ref{eq:lden1}) is given in Ref. \cite{Dhiman2007}.
The equation of motion for baryons, mesons, and photons can be derived from
the Lagrangian density defined in Eq.(\ref{eq:lden1}) by the Euler-Lagrange
Principle \cite{Dhiman2007}.

\begin{table*}
\caption{\label{tab:tab01}The coupling strength parameterizations of the Lagrangian
density defined in Eq.(\ref{eq:lden1}) for BSR3\cite{Dhiman2007} and IOPB I\cite{Kumar2018}
models.  The $\omega$-meson self 
coupling $\zeta$ is equal to 0.0 for BSR3 and 0.0173436 for IOPB-I models, respectively.
The masses of the mesons are taken to be m$_{\omega}=$ 782.5MeV, m$_{\rho}=$ 763MeV,
m$_{\sigma}^{*}=$ 975MeV, and m$_{\phi}=$ 1020MeV for BSR3 parameterizations.
The masses of the mesons are taken to be m$_{\omega}=$ 782.187MeV, m$_{\rho}=$ 762.468MeV
for IOPB-I parameterizations. 
The mass of nucleon, M$_{N}=$939MeV, and the masses of hyperons, M$_{\Lambda}=$ 1116MeV,
M$_{\Sigma}=$ 1193MeV, and M$_{\Xi}=$1313MeV. The values of $\overline{\kappa}$, 
$\overline{\lambda}$, $\overline{\alpha_{1}}$, $\overline{\alpha_{1}}'$, 
$\overline{\alpha_{2}}$, $\overline{\alpha_{2}}'$ and $\overline{\alpha_{3}}'$ 
are multiplied by 10$^2$.}
\begin{tabular}{ccccccccccccc}
\hline 
\hline
EOS&$\bigtriangleup$r&g$_{{\sigma}N}$&g$_{{\omega}N}$&g${_{\rho}N}$&$\overline{\kappa}$&
$\overline{\lambda}$&$\overline{\alpha_{1}}$&
$\overline{\alpha_{1}}'$&$\overline{\alpha_{2}}$&$\overline{\alpha_{2}}'$&
$\overline{\alpha_{3}}'$&m$_{\sigma}$\\
\hline
BSR3&0.20&10.4442&13.5223&13.1170&2.4304&-0.0427&0.1812&0.1597&2.9666&1.2530&0.0972&497.8348\\
IOPB-I&0.221&10.3966&13.3509&11.1257&1.8584&-0.7621&0.0&0.0&0.0&0.0&2.40&500.4870\\
\hline
\hline
\end{tabular}
\end{table*}

The energy density of the uniform matter  within the framework of EFTRMF model is given by;
\begin{equation}
\label{eq:eden}
\begin{split}
{\cal E} & = \sum_{j=B,\ell}\frac{1}{\pi^{2}}\int_{0}^{k_j}k^2\sqrt{k^2+M_{j}^{*2}} dk\\
&+\sum_{B}g_{\omega B}\omega\rho_{B}
+\sum_{B}g_{\rho B}\tau_{3B}\rho
+ \frac{1}{2}m_{\sigma}^2\sigma^2\\
&+\frac{\overline{\kappa}}{6}g_{\sigma N}^3\sigma^3
+\frac{\overline{\lambda}}{24}g_{\sigma N}^4\sigma^4
-\frac{\zeta}{24}g_{\omega N}^4\omega^4\\
&-\frac{\xi}{24}g_{\rho N}^4\rho^4
 - \frac{1}{2} m_{\omega}^2 \omega ^2
-\frac{1}{2} m_{\rho}^2 \rho ^2\\
&-\overline{\alpha_1} g_{\sigma N}
 g_{\omega N}^{2}\sigma \omega^2
 -\frac{1}{2}\overline{\alpha_1}^\prime g_{\sigma N}^2 g_{\omega N}^2\sigma^2 \omega^2\\
 &-\overline{\alpha_2}g_{\sigma N}g_{\rho N}^2 \sigma\rho^2
-\frac{1}{2} \overline{\alpha_2}^\prime g_{\sigma N}^2 g_{\rho N}^2\sigma^2\rho^2\\
 &- \frac{1}{2} \overline{\alpha_3}^\prime g_{\omega N}^2 g_{\rho N}^2
\omega^2\rho^2+\sum_{B}g_{\phi B}\phi\rho_{B}\\
&+\frac{1}{2} m_{\sigma^*}^{2} {\sigma ^*} ^{2}
-\frac{1}{2} m_{\phi}^{2} {\phi} ^{2}.
\end{split}
\end{equation}
The pressure of the uniform matter  is given by
\begin{equation}
\label{eq:pden}
\begin{split}
P & = \sum_{j=B,\ell}\frac{1}{3\pi^{2}}\int_{0}^{k_j}
\frac{k^{4}dk}{\sqrt{k^2+M_{j}^{*2}}} 
- \frac{1}{2}m_{\sigma}^2\sigma^2\\
&-\frac{\overline{\kappa}}{6}g_{\sigma N}^3\sigma^3 
-\frac{\overline{\lambda}}{24}g_{\sigma N}^4\sigma^4
+\frac{\zeta}{24}g_{\omega N}^4\omega^4\\
&+\frac{\xi}{24}g_{\rho N}^4\rho^4
  + \frac{1}{2} m_{\omega}^2 \omega ^2
+\frac{1}{2} m_{\rho}^2 \rho ^2 \\
& +\overline{\alpha_1} g_{\sigma N}
g_{\omega N}^{2}\sigma \omega^2
+\frac{1}{2} \overline{\alpha_1}^\prime g_{\sigma N}^2 g_{\omega N}^2\sigma^2 \omega^2\\
&+\overline{\alpha_2}g_{\sigma N}g_{\rho N}^2 \sigma\rho^2
+\frac{1}{2} \overline{\alpha_2}^\prime g_{\sigma N}^2 g_{\rho N}^2\sigma^2\rho^2\\
&+ \frac{1}{2} \overline{\alpha_3}^\prime g_{\omega N}^2 g_{\rho N}^2
\omega^2\rho^2-\frac{1}{2} m_{\sigma^*}^{2} {\sigma ^*} ^{2}\\
&+\frac{1}{2} m_{\phi}^{2} {\phi} ^{2}.
\end{split}
\end{equation}
The various properties associated with nuclear matter are obtained by using the
parameter sets of Table {\ref{tab:tab01}}. The values of B/A, K, M$^*_{N}$, and 
$\rho_{sat}$ for all these parameter sets lie in a narrow range. B/A=16.11$\pm$0.04MeV, 
K=230.24$\pm$9.80MeV, M$^*_{N}$/M$_{N}$=0.605$\pm$0.004, and $\rho_{sat}$=0.148$\pm$0.003fm$^{-3}$.
\par The composition of nuclear matter (for species i=n, p, $\Lambda$, 
$\Sigma^{-}$, $\Xi^{-}$, $\Xi^{0}$, e$^{-}$, $\mu^{-}$) at fixed baryon 
number density $\rho_{B}$=$\sum_{i}B_{i}\rho_{i}$ are determined in such
a way that the charge neutrality condition,
\begin{equation}
 \sum_{i}q_{i}\rho_{i}=0,
\end{equation}
and the chemical equilibrium conditions
\begin{equation}
\label{eq:mu}
 \mu_{i}=B_{i}\mu_{n}-q_{i}\mu_{e},
\end{equation}
are satisfied.\\
where B$_{i}$ and q$_{i}$ denotes baryon number
and electric charge of the species i.
\subsection{Quarks Phase}
\label{sec:model1}
In the last few years the study of color superconducting phase in quark-gluon plasma has been 
drawing a great interest in discussing
the possible states of quark matter. At QCD perturbative regime the attractive quark
interaction introduces instabilities in the Fermi surface, producing a gap in the quasiparticle
energy spectrum.  
The total thermodynamical potential density of quarks matter may written as \cite{Agrawal2009},
\begin{equation}
\begin{split}
\label{eq:eq1}
 \Omega(\bar{\mu},\mu_{e})&=\Omega_{QQPM}(\bar{\mu})+\Omega(\mu_e)+\Omega_{CFL}({\bar{\mu}}),
 \end{split}
\end{equation}
 where the first term in Eq.(\ref{eq:eq1}) is obtained from the quarks, $\mu_{e}$ is the
 electron chemical potential, so second term is from electrons and third term is from CFL 
 pairing contribution. The contribution to first term 
 in Eq.(\ref{eq:eq1}) is given by \cite{Xia2014}, 

\begin{eqnarray}
\Omega_{QQPM}=-\frac{1}{\pi^2}\sum_{i}\frac{1}{8}\big[\nu_{i}\sqrt{\nu_{i}^2+m_{i}^{*2}}
(2\nu_{i}^2\nonumber\\-3m_{i}^{*2})+3m_{i}^{*4}\ln\large(\frac{\sqrt{\nu_{i}^2
+m_{i}^{*2}}+\nu_{i}}{m_{i}^{*}}\large)\big], 
\end{eqnarray}
The contribution to the second term in Eqn.(\ref{eq:eq1}) is given by,
\begin{eqnarray}
 \Omega({\mu_{e}})=\frac{1}{\pi^{2}}\int_{0}^{\sqrt{\mu_{e}^2-m_{e}^2}}\nu^2
 (\sqrt{\nu^2+m_{e}^2}-\mu_{e})d\nu
\end{eqnarray}

CFL Pairing term contribution is given by,
\begin{equation}
\Omega_{CFL}({\bar{\mu}})=-\frac{3\Delta^{2}\bar{\mu}^{2}}{\pi^{2}},
\end{equation}
where $\Delta$ is CFL color superconducting gap parameter of CFL phase of quark
matter. Medium effects play an important role in describing the properties of quarks
matter via the concept of effective masses.\\
Effective mass is taken to be \cite{Schertler1_1997,Schertler2_1997,Pisarski1989},
\begin{equation}
 m_{i}^{*}=\frac{m_{i0}}{2}+\sqrt{\frac{m_{i0}^2}{4}+\frac{g^2\mu_{i}^2}{6\pi^2}},
\end{equation}
where $m_{i0}$, $\mu_{i}$ and g is the current quark mass, quark chemical potential,
strong interaction coupling constant respectively. 
 In actual practice the strong coupling constant is running \cite{Shirkov1997,Wen2010,
 Patra1996} which is  having a phenomenological expression such as 
 \begin{equation}
   g^2(T,\mu_i)=\frac{48}{29}\pi^2\Big[\ln\Big(\frac{0.8\mu_i^2+15.622T^2}{\Lambda^2}
   \Big)\Big]^{-1},
  \end{equation}
where  $\Lambda$ is the QCD scale-fixing parameter. In the present
calculation, the  value is taken to be 200MeV. In Fig.(\ref{fig01}), variation
in coupling strength parameter for the present case is shown.
It is observed that range of coupling strength parameter is from 3 to 5 for the
chemical potential lying in the range of (510-320)MeV approximately. Also, it 
is showing the decrease in interaction strength with increase in chemical potential
\cite{Stocker1984}. At zero temperature, number densities for all three flavors of 
quarks considered are the same and can be obtained as
 \begin{equation}
  n_i=\frac{d_{q}\nu_{f}^{3}}{6\pi^2}+\frac{2\Delta^2\bar{\mu}}{\pi^2},
 \end{equation} 
 with i = u, d, and s and $\nu_f=\nu_u=\nu_d=\nu_s$.\\
  \begin{equation}
  \nu_f=\Big[\Big(2\mu-\sqrt{\mu^2+\frac{m_{s}^{2}-m_{u}^{2}}{3}}\Big)^2-m_{u}^{2}\Big],
  \end{equation}
  for $\mu$ = $\frac{\mu_{u} + \mu_{d} + \mu_{s}}{3}$ is common Fermi momentum of the
  quark system which depends on the mass of the three quark flavors.\\
 For $m_{u}$ = $m_{d}$ = 0, common fermi momentum becomes,
 \begin{equation}
  \nu=2\mu-\sqrt{\mu^2+\frac{m_{s}^{2}}{3}},
 \end{equation}
For the quarks matter the energy density becomes,
\begin{equation}
\begin{split}
 {\cal E}_{QM}& = \frac{1}{\pi^2}\sum_{i}\frac{3}{8}\big[\nu_{i}\sqrt{\nu_{i}^2+
 m_{i}^{*2}}(2\nu_{i}^2+
  m_{i}^{*2})\\
&-m_{i}^{*4}\ln(\frac{\sqrt{\nu_{i}^2+m_{i}^{*2}}+\nu_{i}}{m_{i}^{*}})\big]+B^{*},\\
\end{split}
\end{equation}
Thus, pressure becomes
\begin{eqnarray}
P_{QP}=-\Omega(\bar{\mu},\mu_{e})-B^{*}.
\end{eqnarray}
B$^{*}$ is the effective Bag function and, can be written as, 
\begin{equation}
  B^{*}=\sum_{i}B_{i}(\mu_{i})+B_{0}.
 \end{equation} 
 The introduction of B$^*$ is done to 
 shows the automatic confinement characteristic in the model, where B$_{0}$ is 
 similar to the conventional MIT bag constant. \\
The $\mu_{i}$-dependent part of effective bag function is calculated as,
\begin{equation}
\label{eq:eq16}
 B_{i}(\mu_{i})=-\int_{m_{i}^*}^{\mu_{i}}\frac{\partial\Omega_{i}}{\partial
 m_{i}^*}\frac{\partial m_{i}^{*}}{\partial\mu_{i}}d\mu_{i}.
\end{equation}
With the above quark mass formulae and thermodynamic treatment, one can get the 
properties of bulk CFL quark matter.\\
In the CFL phase, the three flavors of quarks satisfy the following
conditions:\\
(i) they have equal fermi momenta, which minimizes the free energy of the
system. \\
(ii) they have equal number
densities n$_i$, as a consequence of the first condition, which means that 
n$_i$ = n$_{B}$ and $\mu$$_{i}$ = $\mu$,
for i = {u, d, s}. 
\subsection{Mixed Phase}
We construct the mixed phase of EOS made up of the  hadron matter and  quark
matter by employing the Glendenning construction \cite{Glendenning1992,
Glendenning2000} for hybrid compact star. The equilibrium chemical potential of
the mixed phase corresponding to the intersection of the two surfaces representing
hadron and quark phase can be calculated from the Gibbs condition for mechanical 
and chemical equilibrium at zero temperature 
which reads as,
\begin{equation}
P_{HP}(\mu_e,\mu_n)=P_{QP}(\mu,\mu_e)=P_{MP},
\end{equation}
In mixed phase, we considered chemical equilibrium
at the hadron-quark interface as well as inside each phase \cite{Maruyama2007}, 
so that Eq.(\ref{eq:mu}) implies
\begin{equation}
 \mu_{u}+\mu_{e}=\mu_{d}=\mu_{s},
\end{equation}
\begin{equation}
 \mu_{p}+\mu_{e}=\mu_{n}=\mu_{\Lambda}=\mu_{\Xi^{0}}=\mu_{u}+2\mu_{d},
\end{equation}
\begin{equation}
 \mu_{\Sigma^{-}}+\mu_{p}=2\mu_{n}, 
 \mu_{\Xi^{-}}+\mu_{p}=2\mu_{n}.
\end{equation}
In the mixed phase the local charge neutrality condition is replaced by the 
global charge neutrality which means that both hadron and quark matter are 
allowed to be charged separately.
The condition of the global charge neutrality can be expressed as,  
\begin{equation}
\chi\rho_{c}^{QP}+(1-\chi)\rho_{c}^{HP}=0,
\end{equation}
where, $\chi $ the volume fraction  occupied by quark matter in the mixed 
phase in terms of charge density $\rho_{c}$. The value of the $\chi$ increases
from zero in the pure hadron phase to $\chi=1$ in the pure quark phase.
The energy density ${\cal E}_{MP}$ and the baryon density $\rho_{MP}$ of the 
mixed phase can be calculated as,
\begin{equation}
{\cal E}_{MP}=\chi{\cal E_{QP}}+(1-\chi){\cal E}_{HP}.
\end{equation}
\begin{equation}
\rho_{MP}=\chi\rho_{QP}+(1-\chi)\rho_{HP}.
\end{equation}
With the above evaluated hybrid star EOS presented, we can now analyse the 
structure of the rotating hybrid stars.
\section {Relativistic Rotation of Stars}
The structure of a rapidly rotating CS  is different from the static one,
since the roation can strongly deform the star. We assume CS are steadily rotating 
and have axisymmetric structure. Therefore the 
space-time metric used to model a rotating
star can be expressed as, 
\begin{equation}
 \begin{split}
  ds^{2}=-e^{\gamma+\rho}dt^{2}+e^{2\beta}(dr^2+r^2d\theta^2)\\
  \quad+e^{\gamma-\rho}r^{2}\sin^2{\theta}(d\phi-\omega dt)^2,
 \end{split}
\end{equation}
where the potentials $\gamma$, $\rho$, $\beta$, $\omega$ are functions of r and
$\theta$ only. The matter inside the star is 
approximated by a perfect fluid and the energy-momentum tensor is given by, 
\begin{equation}
 T^{\mu\nu}=({\cal{E}}+P)u^{\mu}u^{\nu}-Pg^{\mu\nu},
\end{equation}
where $\cal{E}$, P and u$^{\mu}$ are the energy density, pressure, and four-velocity, 
respectively. In order to solve Einstein's
field equation for the potentials $\gamma$, $\rho$, $\beta$ and $\omega$, we adopt
the KEH method \cite{KEH1989} and use the RNS code \cite{rns} for calculating
the properties of a rotating star. In Fig.(\ref{fig03}), we present the results of 
the EOSs for the hybrid star and neutron star in the form P=P(${\cal{E}}$). 
\section{Results and Discussions}
\subsection{Equations of State and Non - rotating Hybrid Stars}
In this section, we present the detail results for a set of EOSs of hybrid stars 
in the hadron phase, u, d, s quarks phase and mixed phase. 
In epitome, we employed Baym-Pethick-Sutherland EOS \cite{Baym1971} for low density 
regime from outer crust baryon number density, 
$\rho_b = 6.3\times10^{-12}$fm$^{-3}$ upto the pasta phase $\rho_b = 9.4\times10^{-2}
$fm$^{-3}$ that include the data and description of the neutron drip.
In order to describe the EOS in liquid core of hybrid stars from its inner crust 
$\rho_{crust}$  upto outer core $\rho_b \approx 0.35$fm$^{-3}$,
we used an improved and consistent nuclear matter EOS in $\beta$-equilibrium
based on Extended Field Theoretical Relativistic Mean Field model parameterizations 
BSR3 \cite{Dhiman2007} and IOPB-I \cite{Kumar2018}.
The coupling strength parameters of EFTRMF models for BSR3 and IOPB-I along with 
neutron skin thickness $\Delta r$ of $^{208}$Pb nucleus are presented in 
Table \ref{tab:tab01} of Section-II. These EFTRMF parmeterizations have 
successfully employed to investigate the experimental data of physical 
observables relating to the structural properties of finite nuclei 
\cite{Dhiman2007,Kumar2018} and bulk properties of nuclear matter at
saturation densities.
\par The set of EOSs employed in present work is presented in Table \ref{tab:tab02},
where the, Nucl1 and Nucl2 are the EOSs computed with parameters BSR3 and IOPB-I,
respectively. The EOSs NHy1 and NHy2 are represented by the compositions of nucleons
and hyperons,  where hyperons appeared at a threshhold baryon number density 
$\rho_b \approx 0.35$fm$^{-3}$. In the present work, we have employed the values
of hyperon-meson coupling parameters - 
g$_{mY_{1}}=$ 0.4714,
g$_{mY_{2}}=$ 0.9428,
g$_{{\sigma^*}Y_{1}}=-1.6666$,
g$_{{\sigma^*}Y_{2}}=-1.3333$,
g$_{{\omega^*}Y_{1}}=0.8333$,
g$_{{\omega^*}Y_{2}}=0.4166$,
g$_{{\rho^*}\Lambda}$ = g$_{{\rho^*}\Sigma^{0}}=0$
g$_{{\rho^*}\Sigma^{-}}$ = g$_{{\rho^*}\Sigma^{+}}=1$
g$_{{\rho^*}\Xi^{-}}$ = g$_{{\rho^*}\Xi^{0}}=0.5$
where Y$_{1}$ representing the hyperons as, $\Lambda$, $\Sigma^{-}$, $\Sigma^{+}$,
$\Sigma^{0}$; Y$_{2}$ representing the hyperons $\Xi^{-}$, $\Xi^{0}$ and, m represent 
the $\sigma$ and $\omega$ mesons.
The EOSs presented in Table \ref{tab:tab02} as NmQ1, NmQ2 and NmQ3 are composed of 
nucleons and quarks in $\beta$-equilibrium with 
mixed phase of EOS varies from  $\rho_b \approx$ 0.30-0.40fm$^{-3}$, $\rho_b \approx$
0.34-0.50fm$^{-3}$ and $\rho_b \approx$ 0.39-0.52fm$^{-3}$.
The description for the construction of mixed phase is discussed in Section (IIC).  
Finally, the EOS NHymQ4 has particle composition of nucleons, hyperons and quarks 
in $\beta$ - equilibrium where the $\Lambda$ hyperons are appeared at the threshold 
density of $\rho_b$ $\approx$ 0.35fm$^{-3}$ and the quarks phase 
transition occur at $\rho_b=0.63$fm$^{-3}$. The quarks phase of hybrid EOS has been 
computed by employing Quark Quasiparticle model 
by considering the color superconducting phase with CFL, $\Delta=$50MeV.
Whereas in the present calculation we consider all three quarks masses as,
m$_{u}$=m$_{d}$=4MeV/c$^2$
and m$_{s}$=95MeV/c$^{2}$.
In Table \ref{tab:tab02}, we list our
EOSs and their particles composition, the brief description of 
 theoretical models of dense matter and its parameters, the maximum gravitational 
 mass, $M_G(M_\odot)$ of the compact stars, radius of the compact stars, R$_{max}$,
 baryon number density, $\rho_b$ and energy density,
 ${\cal{E}}$ corresponding to maximum gravitational mass of the non-rotating compact
 stars. Therefore, in the present research work, for set of EOSs the Maximum gravitational
mass of non-rotating hybrid stars varies in the range of (2.38-1.83)$M_\odot$ and 
radius varies as (11.76-12.08)km which satisfies the constraints of the recently
extracted astrophysical observations of gravitational mass and radius of compact 
stars \cite{Most2018,Annala2018,Bose2018,Rezzolla2018}.
\par In Fig.({\ref{fig02}}), we present the variation of theoretically computed pressure 
(MeVfm$^{-3}$) as a function of energy density ${\cal_{E}}$ (MeVfm$^{-3}$) for the EOSs 
employed in the present work. In the upper panel, we present the EOSs Nucl1, Nucl2, 
NmQ1, NmQ2 and NmQ3 represented by black solid curve, brown dotted curve, orange 
double-dash-dotted curve, red small-dashed curve and indigo long dashed dotted curve 
respectively. The open coloured circles represent the boundaries of mixed phase consisting 
of nucleons and quarks in $\beta$-equilibrium. Further, the pressure at densities 
$2\rho_0$ and $6\rho_0$ of Nucl1 EOS are $5.25\times10^{34}dyne/cm^2$ and $9.648\times10^{35}
dyne/cm^2$, respectively, are comparable with the recently measured pressure at same 
densities \cite{Abbott2018}. However, in the present work the magnitude of pressure is 
decreasing at $6\rho_0$, as the EOSs become softer with the appearance of 
hyperons and quarks matter phases and, the pressure at $6\rho_0$ for EOS of NHymQ4 model 
becomes $4.2\times10^{35} dyne/cm^2$.
\par In Fig.(\ref{fig03}), we present the results for relationship between gravitational mass
and radius of non-rotating compact star for various EOSs constructed in the present work. The 
region excluded by causality, light green solid line and rotation constraints of neutron star 
XTE J1739-285 solid maroon line are given. The mass and radius 
 limit estimated from Vela pulsar glitches $\Delta I/I$=0.014 is shown as blue solid line. 
 The mass limits of pulsars PSR J1614-2230 and PSR J0348+0432 are plotted for comparison. 
The limits on compact star mass and radius from Ozel's analysis of EXO 0748-676 with 1$\sigma$ 
(dark solid black line) and 2$\sigma$, extended black line error bars are also shown. 
The mass radius relationship obtained from the EOS by extracted data of EOS by using 
QMC+Model A is also shown. The orange region bounded by the dotted maroon lines is representing the 
mass - radius relationship extracted for the proposed pulsar PSRJ0437-4715 for 3$\sigma$ 
confidence level in the NICER program \cite{Gendreau2012}. The plausible set of EOSs used
in the present calculation have bounds for maximum gravitational mass, 
1.83$M_{\odot}$$\le$M$_{G}$$\le$2.38M$_{\odot}$ and stars's radius, 11.76km-12.08km 
as shown in Table \ref{tab:tab01}, these limits of mass and radii are well within the 
recenlty extracted limits on mass and radii \cite{Most2018,Annala2018,Bose2018}.
\subsection{Rotating Hybrid Stars}
\par The rotational frequency is a directly measurable physical quantity of the pulsars, 
and the Keplerian (mass-shedding) frequency f$_K$ is one of the most studied physical 
quantities for rotating stars \cite{Zacchi2016,Alford2015,Lattimer2007,Benhar2005}.
In Table \ref{tab:tab03}, we present the structural properties of rotating compact stars 
at keplerain frequency. We present the values of maximum gravitational mass 
$M_{max}(M_\odot)$, its corresponding equatorial radius $R_{max}$(km), central 
energy density ${\cal{E}}_{c}(\times10^{15}g/cm^{-3})$, baryon density $\rho$($fm^{-3}$), 
the maximum Keplerian frequency $f_{K}$(Hz) and the empirical approximation 
of maximum frequency f$_{max}$. The empirical formula \cite{Haensel2016} of
f$_{max}$ is the correspondence between two extremal configurations of the static 
configuration with a maximum allowable mass, M$_{max}^{stat}$, R$^{stat}_{{M_{max}}}$, 
and stably rotating configuration with a maximum allowed frequency, can be written as,
\begin{equation}
\label{eq:fmax}
 f_{max}=f_{0}\Big(\frac{M_{max}^{stat}}{M_{\odot}}\Big)^{\frac{1}{2}}\Big(\frac{ 
 {R_{M_{max}}^{stat}}}{10km}\Big)^{(-\frac{3}{2})},
\end{equation}
where M$^{stat}_{max}$ and R$_{M_{max}}$ are the maximum gravitational mass and radius of
the static configuration, respectively. The f$_{0}$ is a constant frequency equals to 1.22kHz,
which does not depend on the theoretical model of EOS. 
This formula for f$_{max}$ is authentic for all CSs composition of hadrons in $\beta$-
equilibrium, self-bound quarks matter stars and the hybrid compact stars. 
The Fig.(\ref{fig04}), represent the theoretical results for the variation of gravitational mass as 
a function of equatorial radius for various model of EOSs. In Fig.({\ref{fig05}}), we
presented keplerian frequency as a function of gravitational mass for EOSs used in the 
present work. It can be observed from Fig.(\ref{fig05}) that the keplerian frequency increases 
monotonically  both for hadronic star as well as hybrid star as a function of gravitational 
mass (M$_{G}$). 
The maximum gravitational mass and keplerian frequencies of the hybrid compact stars 
decreases as the hyperons and quarks appears in core of star and, consequently, the size of 
hybrid stars increases. 
\par In order to visualize better the complex relations among M, R, and f$_{K}$, we present
in Fig.(\ref{fig06}) gravitational mass as a function of the equatorial radius at various 
fixed rotation frequencies with EOS NHymQ4. The stable configurations are constrained 
by the kepler and SAI conditions at large and small radius, respectively.
At a low frequency 465.1Hz, the lower boundary of mass M is fixed by the kepler 
condition and upper boundary by SAI condition. As the frequency of the known pulsars 
increases, f=465.1Hz \cite{Freire2011}, 641.9 Hz \cite{Friedman1988}, 716.4Hz 
\cite{Hessels2006}, 1122Hz \cite {Kaaret2007}, 
the SAI mark point moves more and more to the left side of the mathematical 
maximum. As the frequency increases further to f=1122Hz, the lower
(upper) boundary values of mass M are fixed by the SAI
(kepler) conditions. Also, the gravitational masses of the sequences as a function of 
equatorial radius for static, keplerian and evolutionary sequences is shown along with the
evolutionary sequences.
\subsection{Phase transition caused by rotational evolution}
The possibility of a phase transition to quarks phase which is caused by the rotational 
evolution have been widely discussed in the literature \cite{Marczenko2018,Wei2017,Haensel2016,
Ayvazyan2013,Zdunik2006}. For a constant 
baryonic mass, a rotating star loses its 
rotational energy by magnetic dipole radiation, which makes the star spin-down and
the central density increases. When the central density of a star reaches a critical 
value, the phase transition from the hadronic matter to quarks phase will take place,
and star converts to a hybrid star. As the star continues spinning down and the central
density continues increasing, more and more quarks matter phase appears in the core of 
the hybrid star. The Fig.(\ref{fig07}) depicts the profile of the pressure as a function 
of equatorial distance of a compact star, whose rest mass is  2.226M$_{\odot}$,  on the 
internal radius of four spin frequencies in the range as, 0$\le$$\nu$$\le$716.4Hz.
For the frequency of fastest rotating stars 716Hz \cite{Hessels2006}, it is found that
the hybrid star do not contain quarks matter inside the core and it is composed of 
neutrons, protons and hyperons only. The stellar radius shrinks along with the regions
of particles compositions of hyperons and mixed phase with the spin down frequencies and, 
however, the region of neutrons and protons expanded and quarks matter core appears at 
the frequency 465.1 Hz of pulsar \cite{Freire2011}. 
At asymptotically slow rotation rates, the star represents a dense CFL quark phase that
extends upto the radius 2.3km and, is surrounded by a mixed phase followed by hyperonic 
and nucleonic phase in the region between 2.3km$\le$R$\le$12.07km.
Fig.(\ref{fig07}) gives a true reflection of the internal structure of the stars as it 
spins-up or down, if the dynamical timescales are much larger than the timescales required
for the nucleation of the CFL quarks matter phase.
\subsection{The ms Pulsars and Limits of Radii}
Recently, many attempts have been made to constrain the EOSs model with the extracted radii \cite{Gendre2003,Bhattacharyya2005,Bogdanov2007,Ozel2016} 
of compact stars from astrophysical observations by making use of various spectroscopic 
and timing methods. In this work we have attempted to predict the limits of radii of ms 
pulsars displayed in Table \ref{tab:tab04} by employing the theoretical models NHymQ4 of 
EOS for hybrid star presented Table \ref{tab:tab02}.
In Table \ref{tab:tab04}, we present the catalogue of the recently observed pulsar's 
precise gravitational masses and their numerical computed spinning frequency in the 
range of 2.15ms-5.76ms. This knowledge of astrophysical data for mass and spin 
values  of the pulsars may be enough to understand the behaviour of other compact stars, 
including some of the fast spinning accreting stars in LMXBs. In Table \ref{tab:tab05}, 
the theoretically computed limits for the results of spin-down sequences of hybrid star 
for gravitational mass M$_{G}(M_{\odot})$, equatorial radii R$_{eq}$(km), 
baryonic mass M$_{B}$, redshift Z, moment of inertia I, the ratio of rotational to
gravitational energies T/W, dimensionless angular momentum cJ/GM$_{G}$ $^2$ and ratio
of equatorial radii at pole to equator R$_p$/R$_e$ are presented. 
 The corresponding values of central energy density e$_{c}(\times 10^{15}$gcm$^{-3})$ and 
 baryon number density $\rho_c$($\times$10$^{14}$gcm$^{-3}$) are also presented in the Table
 \ref{tab:tab05}. These limits for results have been obtained corresponding to the observed 
 rotating frequencies $\nu$
at 465.1Hz, 416.7Hz, 366.0Hz, 339.0Hz, 317.5Hz, 315.5Hz, 287.4Hz, 279.3Hz and 173.6Hz 
corresponding to 
the pulsars shown in Table \ref{tab:tab04},  PSR J1903+0327\cite{Champion2008,Freire2011}, 
PSR J2043+1711\cite{Demorest2010,Fonseca2016}, 
PSR J0337+1715\cite{Ransom2014}, PSR J1909-3744\cite{Jacoby2003,Desvignes2016}, PSR J1614-2230\cite{Demorest2010,Fonseca2016},  
PSR J1946+3417\cite{Barr2013,Ozel2016}, PSR J0751+1807\cite{Lundgren1993,Jacoby2003}, PSR J2234+0611\cite{Deneva2013,Ozel2016} and 
PSR J0437-4715\cite{Johnston1993,Reardon2015}. 

The densest nuclear matter in the universe possibly nestles in the core of compact stars. 
Therefore, it is inveigling to know how dense matter can be for observed in ms pulsars with
measured masses presented Table \ref{tab:tab04}.
These investigations will also characterise the compact star populations,
and help us to understand their particles composition, formation and evolution. 
In view of recent astrophysical observations of pulsars with gravitational mass more than
1.4M$_\odot$, the Fig.\ref{fig08}, presents the theoretical 
limits of the radii of various well known millisecond pulsars presented in Table \ref{tab:tab04} 
as a function of their rotating frequencies. The evolutionary sequences of the pulsars, 
theoretical limits of the radii and the gravitational mass of these pulsars have been computed 
by using NHymQ4 equation of state of hybrid star.
For slow rotating hybrid star of gravitational maximum mass around 1.4M$_\odot$, the limits
of the radii varies as 13.71km$\leq$R${_{eq}}\leq$14.10km which is very close with the 
recently extracted bounds on radii of the compact stars \cite{Lattimer2014,Bhattacharyya2017}. 
Whereas for the spinning hybrid star with gravitational mass around 2.0M$_\odot$, we predict the
limits of radii as, 12.85km$\leq$R${_{eq}}\leq$13.50km. 

\subsection{Static versus spinning configurations}
In Table \ref{tab:tab06}, we present the theoretically computed radii for the ms pulsars 
given in Table \ref{tab:tab04} for non-spinning configurations, and the corresponding
percentage bias in the allowed EOS  of hybrid stars for model NHymQ4, if this model of 
EOS is constrained using an accurate $\pm$5$\%$ for equatorial radius measurement.
For constraining EOS models using the radius measurements of radio ms pulsars, for
example with NICER measurements, it is commonly argued that the comparably slow spin 
rates of compact stars nominally affect the radius. Nonetheless, the spin-related 
systematic error depends on the accuracy desired to measure the equatorial radius,
R$_{eq}$ with desired accuracy of ($\xi$), as recommended about 5$\%$ \cite{Lattimer2001}.
NICER can also measure the stellar radius with a similar accuracy \cite{Gendreau2012}.
Therefore, we compute the radius R of ms pulsars for NHymQ4 EOS model for 
static configuration, and calculate the percentage difference $\eta$ between R and
R$_{eq}$ using the gravitational radii R for static configurations. If the EOS model 
is constrained using a R$_{eq}$ result  measured with a percentage accuracy of $\pm$ $\xi$ 
and using theoretical non-spinning configuration, then there can be
($\eta$/2$\xi$ ) $\times$ 100$\%$ of bias in the allowed EOS model. 
This means, $\sim$($\eta$/2$\xi$ ) $\times$ 100$\%$ of the allowed EOS 
models would be falsely allowed because of the systematic difference
between R and R$_{eq}$. It can be noticed from the Table \ref{tab:tab06}
that the bias is quite high for the faster spinning pulsars for selected EOS
model NHymQ4, whereas, the magnitude of the bias is smaller for slowly spinning
pulsars. The computation of a bias will be indispensable to know the authenticity
of constraints on EOS models. However, for a given $\eta$ value, the $\xi$ is 
expected to decrease with the availability of better instruments in the future. 
This will increase the bias values shown in Table \ref{tab:tab06}, which shows
the usefulness of the first extensive tabulation of these values and the 
computation of spinning configurations in the present work.
\subsection{Tidal deformability}
The dimensionless love
number k$_2$ is an important physical quantity to measure the internal 
structure of the constituent body. The dimensionless tidal deformability
parameter  $\Lambda$ depends on the neutron star compactness C and a dimensionless
quantity k$_{2}$ as, $\Lambda$=2k$_{2}$/${3C^5}$.
To measure the love number k$_{2}$ along with the evaluation of the TOV 
equations we have to compute y=y(R) with 
initial boundary condition y(0)=2 from the first-order
differential equation \cite{Hinderer2008,Hinderer2009,Hinderer2010,Damour2010}
 iteratively, 
 \begin{eqnarray}
 \label{y}
   y^{\prime}=\frac{1}{r}[-r^2Q-ye^{\lambda}\{1+4\pi Gr^2(P-{\cal{E}})\}-y^{2}], 
 \end{eqnarray}
where  Q$\equiv$4$\pi$Ge$^{\lambda}$(5${\cal{E}}$+9P+$\frac{{\cal{E}}+P}{c_{s}^2})$
 -6$\frac{e^{\lambda}}{r^2}$-$\nu^{\prime^2}$ and       
  e$^{\lambda} \equiv(1-\frac{2 G m}{r})^{-1}$ and, $\nu^{\prime}\equiv$ 2 G e$^{\lambda}$ 
       ($\frac{m+4 \pi P r^3}{r^2}$).
First, we get the solutions of Eq.(\ref{y}) with boundary condition, y=y$_{2}$(R),
then the electric tidal Love
numbers k$_{2}$ is calculated from the expression as,
\begin{eqnarray}
 k_{2}=\frac{8}{5}(1-2C)^{2}[2C(y_{2}-1)-y_{2}+2]\{2C(4(y_{2}+1)C^4\nonumber\\
 +(6y_{2}-4)C^{3}
 +(26-22y_{2})C^2+3(5y_{2}-8)C-3y_{2}+6)\nonumber\\
 -3(1-2C)^2
 \quad(2C(y{_2}-1)-y_{2}+2) \log(\frac{1}{1-2C})\}^{-1}.\nonumber\\
\end{eqnarray}
\par Fig.(\ref{fig09}) right panel presents the dimensionless tidal love number k$_2$ as
a function of gravitational mass of the compact star. The value of k$_2$ suddenly
decreases with increasing gravitational mass from 0.95$M_\odot$ onwards. The value of k$_2$
is low at higher and lower gravilational masses of the compact star and, indicating that 
the quadrupole deformation is maximum for intermediate ranges of masses.
Fig.(\ref{fig09}) left panel presents the dimensionless tidal deformability as a
function of gravitational mass for the CS with selected eight EOSs in the
present work. It is found that dimensionless tidal deformability decreases with increase
in gravitational mass of compact stars. The obtained value of $\Lambda_{1.4}$ for 
1.4 $M_\odot$ for selected EOSs is from 606$M_\odot$ to 656$M_\odot$, 
which is consistent with recent constraint proposed for tidal deformability
$\Lambda_{1.4}$ from the GW170817 event \cite{Abbott2017,Abbott2018}
for equation of state of dense nuclear matter.
Finally, in Fig.(\ref{fig10}) we plot the tidal deformability parameters $\Lambda_{1}$ 
and $\Lambda_{2}$ which have relationship with the neutron star
binary companion having a high mass M$_{1}$ and a low
mass M$_{2}$, repectively, associated with GW170817 event \cite{Abbott2019}. 
The dark green band and other colored solid line curves represents the tidal parameters 
for the representative set of EOSs (see Table \ref{tab:tab02}) 
and other EOS model shown in Fig.({\ref{fig10}), respectively. The curves of
tidal parameters are ending at   $\Lambda_{1} 
=\Lambda_{2}$ boundary. The tidal parameters from RMF models IUFSU, TM1, and G2 are
calculated by using our computer code and, the curves corresponding to 
the APR4 and SLy models are extracted from Fig.(1) \cite{Abbott2018} for comparison.
The tidal parameters results are obtained by varying the high mass M$_1$
independently in the range as, 1.36 $\leq$M/M$_{\odot}$ $\leq$1.61 and,
obtained the low mass  partner M$_{2}$ of the neutron star merger by keeping the chirp 
mass M$_{chirp}$ = (M$_{1}$ M$_{2}$)$^{(3/5)}$/(M$_{1}$ +M$_{2}$)$^{1/5}$ fixed at the 
observed value, 1.186M$_{\odot}$ \cite{Abbott2019}. The low mass M$_2$ of the neutron 
binary is obtained in the range 1.16 $\leq$ M/M$_{\odot}\leq$1.36, as presented in the
Table \ref{tab:tab07} for NHymQ4 EOS model. The long dashed black lines signifies the
50 $\%$ and the 90$\%$ probabilities of credible regions for waveform model TaylorF2 
for low-spin priors \cite{Abbott2019}. 
Recently, LIGO and VIRGO detectors precisely measured the value of $\tilde{\Lambda}$ of 
the BNS 
\cite{Abbott2017}.
The weighted dimensionless tidal deformability $\tilde{\Lambda}$  of the BNS  of masses M$_{1}$ and M$_{1}$, and  $\tilde{\Lambda}$ = $\Lambda_1$ = $\Lambda_2$ is 
defined \cite{Abbott2017,Favata2014} as,
\begin{eqnarray}
 \tilde{\Lambda}=\frac{16}{13}\Bigg(\frac{(M_1+12M_2)M_1^4}{(M_1+M_2)^5}\Lambda_1 + 
 \frac{(M_2+12M_1)M_2^4}{(M_1+M_2)^5}\Lambda_2\Bigg),\nonumber\\
\end{eqnarray}
with weighted tidal correction $\Delta\tilde{\Lambda}$  \cite{Wade2014},
\begin{eqnarray}
\Delta\tilde{\Lambda}=\frac{1}{2}\Big[\sqrt{1-4\eta}\Big(1-\frac{13272}{1319}\eta+
\frac{8944}{1319}\eta^2\Big)(\Lambda_1+\Lambda_2)+\nonumber\\
\Big(1-\frac{15910}{1319}\eta+ \frac{32850}{1319}\eta^2+\frac{3380}{1319}\eta^3\Big)
(\Lambda_1-\Lambda_2)\Big].\nonumber\\
\end{eqnarray}
Where $\eta$ = M$_1$M$_2$/M$^2$ is the symmetric mass ratio and M = M$_1$ +M$_2$ is 
the total mass. The tidal parameters $\Lambda_1$ and $\Lambda_2$ satisfy the condition
that the M$_1 \geq$ M$_2$. In Table \ref{tab:tab07}, we present the BNS
masses M$_{1}$(M$_{\odot}$),$M_{2}$(M$_{\odot}$) and their corresponding radii R$_{1}$,
R$_{2}$ in km, dimensionless tidal deformability parameters ($\Lambda_{1}$, $\Lambda_{2}$),
weighted dimensionless tidal deformability $\tilde{\Lambda}$, tidal correction
$\Delta\tilde{\Lambda}$ and chirp radius R$_{c}$ in km for EOS model NHymQ4.
Radius of chirp mass is 2M$_{chirp}\tilde{\Lambda}^{1/5}$.
It is noticed that the values of $\tilde{\Lambda}\leq$ 800 in the low-spin priors,  
which is very well consistent with the recent observations \cite{Abbott2017}.
The weighted tidal deformation is found to be lie in the range 689$\leq$ 
$\Delta\tilde{\Lambda}\leq$732 and the chirp 
radius is in the range 8.77 km $\leq$R$_c$ $\leq$ 8.87 km
for equal and unequal-mass binary neutron stars, as shown in Table \ref{tab:tab07}.
\section{Conclusions}
In the present research work, we have obtained a plausible set of hybrid equations of
state for superdense hadron-quarks matter which satisfies the constraints provided by 
finite nuclei, bulk nuclear matter, the observational data of frequencies and 
maximum gravitational mass 2$M_\odot$ with their extracted radii, keplerian limits,
secular axisymmetric instability and limits of tidal deformabilities from GW170817 
BNS merger. The set of EOSs is constructed with the EFTRMF
approach for hadronic matter and the QQPM model for quark matter, and assuming 
the phase transition under the Gibbs construction. Further, we considered in 
region of baryons densities as, $0 \leq \rho \leq 2\rho_0$, the EOS is composed of
neutrons and protons in beta equilibrium, $2\rho \leq \rho \leq 4\rho_0$ hyperons 
appears and $4\rho_{0} \leq \rho \leq 1.6fm^{-3}$ the quarks matter appears in
$\beta$-equilibrium. The set of EOSs is used to determine the maximum
gravitational mass, equatorial radius, rotation frequency of stable stellar
configuration and tidal deformability of hybrid stars. Further, the hybrid EOS NHymQ4
model has been employed to investigate
the phase transition cuased by the rotational evolution, the ms pulsars and limits of
their radii and tidal deformability of hybrid CS.
\par We investigated the phase transition induced by the
spin-down of pulsars with a constant baryonic mass 2.226M$_{\odot}$  on the 
internal radius of four spin frequencies in the range as, 0$\le\nu\le$716.4Hz.
It is found that the gravitational radius, regions of hadrons and mix phase of 
hybrid star shrinks with spin down frequencies and, the regions of nucleons matter
and quarks matter expanded along the equatorial distance of the hybrid stars. 
 However, for the frequency of fastest rotating stars 716Hz, the hybrid star do 
 not contain quarks matter inside the core and it is composed of neutrons, protons
 and hyperons only and quarks matter core appears at the frequency 465.1 Hz of hybrid
 star. We computed and investigated the catalogue of the recently observed pulsars in
 terms of their structural properties and the results of spin-down sequences of hybrid
 star for gravitational mass M$_{G}(M_{\odot})$, equatorial radii R$_{eq}$(km), 
baryonic mass M$_{B}$, redshift Z, moment of inertia I, ratio of rotational 
to gravitational energies T/W, dimensionless angular momentum cJ/GM$_{G}$ $^2$ and
ratio of equatorial radii at pole to equator R$_p$/R$_e$. 
In case of recently observed  pulsars with gravitational mass more than 1.4M$_\odot$, 
we have extracted the theoretical limits of the radii of various well known millisecond
pulsars presented in Table \ref{tab:tab04} as a function of their rotating frequencies 
by using NHymQ4 equation of state of hybrid star.
For slowing rotating hybrid star of gravitational maximum mass around 1.4M$_\odot$, 
the limits of the radii varies as 13.71km$\leq$R${_{eq}}\leq$14.10km which is very 
close with the recently extracted bounds on radii of the compact stars 
\cite{Lattimer2014,Bhattacharyya2017}. Whereas for the spinning hybrid star with
gravitational mass around 2.0M$_\odot$, we predict the limits of radii as, 
12.85km$\leq$R${_{eq}}\leq$13.50km.  

\par In the last we estimated love number and tidal deformability for the set of EOSs.
The tidal deformability parameters do not differ much from each other for set of EOSs
Fig.(\ref{fig10}) and lie within the acceptable range as, $245 \leq \Lambda_1 \leq 732$ and
$732 \leq \Lambda_2 \leq 1828 $ and consistent with the associated components neutron binary
stars of GW170817 event \cite{Abbott2019}. Further, by investigating the consequences of
dimensionless tidal deformability $\Lambda_{1.4} \leq 800$  
provided by LIGO-Virgo collaboration, we extract a limit of the stellar radius of neutron
star of mass 1.4M$_\odot$, R$_{1.4} \leq 13.24$km.
The limits of radii suggested for binary neutron stars 
 R$_{1}$=11.9$^{+1.4}_{-1.4}$ and R$_{2}$ =11.9$^{+1.4}_{-1.4}$
 from GW170817 event \cite{Abbott2019} implies that the EOS of dense nuclear matter 
 at high densities is soft and, the 
 evolution from stiff to soft EOS may indicate the phase transition in the interior 
 of neutron star.
\pagebreak
\begin{table*}
 \caption{\label{tab:tab02}The equations of state and its particles composition, 
 the brief description theoretical models of dense matter and its parameters, 
 the maximum gravitational mass, $M_G(M_\odot)$ of the compact stars, 
 maximum radius of the compact stars, R$_{max}$ and energy density, $\cal{E}$ 
 corresponding to maximum gravitational mass of the nonrotating compact stars.
 The data of EOSs are available with authors through email.}
 \begin{tabular}{ccccccc}
\hline
 \multirow{1}{*}{} &\multirow{1}{*}{}&\multirow{1}{*}{Composition}
 &\multicolumn{1}{c}{Maximum non -}&\\
 \multirow{1}{*}{No.} &\multirow{1}{*}{EOS}&\multirow{1}{*}
 { and Brief description}&\multicolumn{1}{c}
 {spinning mass}&\multicolumn{1}{c}{R$_{max}$(km)}&\multicolumn{1}{c}
 {${\cal{E}}(\times10^{15}g/cm^{-3})$}\\
   \multirow{1}{*}{} &\multirow{1}{*}{}&\multirow{1}{*}{Model}
   &\multicolumn{1}{c}{(M$_{\odot}$)}&\\
 \hline
1 &Nucl1&n,p; BSR3&2.38&12.02&1.955\\
2 &Nucl2&n,p;IOPB-I &2.16&12.07&1.92\\
3 &NHy1&n, p, $\Lambda$,$\Sigma$,$\Xi$; BSR3&1.99&11.58&2.196\\
4 &NHy2&n, p, $\Lambda$,$\Sigma$,$\Xi$,&1.83&11.76&2.08\\
&&IOPB-I&\\
5&NmQ1&n, p, u, d, s&1.93&11.90&2.028\\
&&$\Delta$ = 50MeV,BSR3&\\
&&(B$_{0}$)$^{1/4}$=150MeV\\
6&NmQ2&n, p, u, d, s&1.89&12.0&1.978\\
&&$\Delta$ = 50MeV,BSR3&\\
&&(B$_{0}$)$^{1/4}$=155MeV\\
7&NmQ3&n, p, u, d, s&1.85&11.75&2.07\\
&&$\Delta$ = 50MeV,IOPB-I&\\
&&(B$_{0}$)$^{1/4}$=155MeV\\
8&NHymQ4&n, p, $\Lambda$,$\Sigma$,$\Xi$ &\\
&&+u,d,s quarks &1.95&12.08&1.945\\
&& $\Delta$ = 50MeV&\\
&&(B$_{0}$)$^{1/4}$=150MeV, BSR3\\
\hline
\end{tabular}
\end{table*}
\begin{table*}
 \caption{\label{tab:tab03} The structural properties of rotating compact
 stars, the maximum gravitational mass $M_{max}(M_\odot)$
 and its corresponding equatorial radius $R_{max}$(km), central
 energy density ${\cal{E}}_{c}(\times10^{15}g/cm^{-3})$, the maximum 
 Keplerian frequency $f_{K}$(Hz), the approximate 
 value of maximum frequency f$_{max}$ as defined in Eq.(\ref{eq:fmax}).}
\begin{tabular}{|cc|ccccccccc|}
\hline 
 &&&Nucl1&Nucl2&NHy1&NHy2&NmQ1&NmQ2&NmQ3&NHymQ4\\
 \hline
&&${M_{max}}/M_{\odot}$&2.87&2.60&2.39&2.20&2.32&2.28&2.52&2.37\\
&Keplerian&R$_{max}$(km)&16.06&16.26&16.22&16.57&16.42&16.95&16.73&16.75\\
&&${\cal{E}}_{c}(\times10^{15}g/cm^{-3})$&1.67&1.70&1.75&1.67&1.72&1.57&1.53&1.61\\
&&f$_{K}$(Hz)&1488&1396&1352&1261&1313&1245&1309&1288\\
&&f$_{max}(Hz)$&1428&1352&1381&1294&1306&1273&1303&1283\\
\hline
\end{tabular}
\end{table*}
\begin{table*}
 \centering
 \caption{\label{tab:tab04}List of ms pulsars with measured gravitational 
 mass and less than 10 ms spin-period}
 \begin{tabular}{|c|c|c|c|c|}
\hline
\multirow{1}{*}{No.} &\multirow{1}{*}{Pulsar}&\multirow{1}{*}{Spin-period 
[frequency]}&\multicolumn{1}{c|}{Mass}&\multicolumn{1}{c|}{References}\\
 \multirow{1}{*}{} &\multirow{1}{*}{name}&\multirow{1}{*}{(ms [Hz])}
 &\multicolumn{1}{c|}{(M$_{\odot}$)}&\multicolumn{1}{c|}{}\\
 \hline
1 &J1903+0327&2.15 [465.1]&1.667$^{+0.021}_{-0.021}$&\cite{Champion2008,Freire2011}\\
2 &J2043+1711&2.40 [416.7]&1.410$^{+0.21}_{-0.18}$&\cite{Abdo2010,Guillemot2012,Fonseca2016}\\
3 &J0337+1715&2.73 [366.0]&1.4378$^{+0.0013}_{-0.0013}$&\cite{Ransom2014}\\
4 &J1909-3744&2.95[339.0]&1.540$^{+0.027}_{-0.027}$&\cite{Jacoby2003,Desvignes2016}\\
5&J1614-2230&3.15 [317.5]&1.928$^{+0.017}_{-0.017}$&\cite{Demorest2010,Fonseca2016}\\
6&J1946+3417&3.17 [315.5]&1.832$^{+0.028}_{-0.028}$&\cite{Barr2013,Ozel2016}\\
7&J0751+1807&3.48 [287.4]&1.640$^{+0.15}_{-0.15}$&\cite{Lundgren1993,Jacoby2003}\\
8&J2234+0611&3.58 [279.3]&1.393$^{+0.013}_{-0.013}$&\cite{Deneva2013,Ozel2016}\\
9&J0437-4715&5.76 [173.6]&1.440$^{+0.07}_{-0.07}$&\cite{Johnston1993,Reardon2015}\\
\hline
 \end{tabular}
\end{table*}
\begin{table*}
 \centering
 \caption{\label{tab:tab05} The theoretically computed limits for the 
 results of spin down sequences of hybrid star for gravitational 
 mass M$_{G}(M_{\odot})$, equatorial radii R$_{eq}$(km), baryon mass M$_{B}$, 
 redshift Z, moment of inertia, the ratio of rotational to gravitational energies 
 T/W, dimensionless angular momentum cJ/GM$_{G}$ $^2$ and ratio of equatorial radii
 at pole to equator. These limits for results have been obtained corresponding to
 the observed rotating frequencies $\nu$ at 465.1Hz, 416.7Hz, 366.0Hz, 339.0Hz, 317.5Hz,
 315.5Hz, 287.4Hz, 279.3Hz and 173.6Hz
corresponding to pulsars  PSR J1903+0327\cite{Champion2008,Freire2011},
PSR J2043+1711\cite{Demorest2010,Fonseca2016},PSR J0337+1715\cite{Ransom2014},
PSR J1909-3744\cite{Jacoby2003,Desvignes2016}, PSR J1614-2230\cite{Demorest2010,Fonseca2016},PSR J1946+3417\cite{Barr2013,Ozel2016},
 PSR J0751+1807\cite{Lundgren1993,Jacoby2003}, PSR J2234+0611\cite{Deneva2013,Ozel2016} andPSR J0437-4715\cite{Johnston1993,Reardon2015}. 
 The corresponding values of central energy density 
 e$_{c}=(\times 10^{15}gcm^{-3}$) and baryon number density $\rho_c$($\times$10$^{14}$gcm$^{-3}$)
 are also shown in the table.}
\begin{tabular}{|c|c|c|c|c|c|c|c|c|c|c|}
\hline
\multicolumn{9}{|c|}{PSR J1903+0327, $\nu$=465.1Hz, M$_{G}$=1.667$^{+0.021}_{-0.021}$} \\
\hline
$\rho_c$&e$_{c}$&M$_{B}$($M_{\odot}$)&R$_{eq}$(km)&Z&I($\times$10$^{45}$gcm$^2$)&
T/W&cJ/GM$_{G}$ $^2$&R$_{p}$/R$_{e}$\\
\hline
6.50&0.75$^{+0.02}_{-0.02}$&1.84$^{+0.03}_{-0.03}$&14.07$^{-0.03}_{+0.03}$&0.26$^{+0.00}_{-0.00}$&2.39$^{+0.03}_{-0.03}$&0.02&0.79$^{+0.01}_{-0.01}$&0.91\\
\hline
\multicolumn{9}{|c|}{PSR J2043+1711, $\nu$=416.7Hz, M$_{G}$=1.41$^{+0.21}_{-0.18}$} \\
\hline
5.36&0.60$^{+0.12}_{-0.06}$&1.53$^{+0.26}_{-0.21}$&14.12$^{-0.12}_{+0.08}$&0.20$^{+0.04}_{-0.03}$&1.90$^{+0.38}_{-0.32}$&0.02&0.57$^{+0.11}_{-0.09}$&0.91\\
\hline
\multicolumn{9}{|c|}{PSR J0337+1715, $\nu$=366.0Hz, M$_{G}$=1.4378$^{+0.0013}_{-0.0013}$} \\
\hline
5.53&0.61$^{+0.00}_{-0.00}$&1.57$^{+0.00}_{-0.00}$&13.99$^{-0.00}_{+0.00}$&0.21$^{+0.00}_{-0.00}$&
1.93$^{+0.00}_{-0.00}$&0.01&0.51$^{+0.00}_{-0.00}$&0.93\\
\hline
\multicolumn{9}{|c|}{PSR J1909-3744, $\nu$=339.0Hz, M$_{G}$=1.540$^{+0.027}_{-0.027}$} \\
\hline
6.0&0.67$^{+0.02}_{-0.02}$&1.69$^{+0.03}_{-0.04}$&13.90$^{-0.01}_{+0.01}$&0.23$^{+0.01}_{-0.01}$&
2.11$^{+0.05}_{-0.05}$&0.01&0.51$^{+0.01}_{-0.01}$&0.94\\
\hline
\multicolumn{9}{|c|}{PSR J1614-2230, $\nu$=317.5Hz, M$_{G}$=1.928$^{+0.017}_{-0.017}$} \\
\hline
10.56&1.30$^{+0.09}_{-0.07}$&2.18$^{+0.02}_{-0.02}$&12.98$^{-0.13}_{+0.11}$&0.34$^{+0.01}_{-0.01}$&
2.53$^{-0.02}_{+0.01}$&0.01&0.57$^{-0.00}_{+0.00}$&0.96\\
\hline
\multicolumn{9}{|c|}{PSR J1946+3417, $\nu$=315.5Hz, M$_{G}$=1.832$^{+0.028}_{-0.028}$} \\
\hline
8.61&1.03$^{+0.06}_{-0.05}$&2.06$^{+0.04}_{-0.04}$&13.43$^{-0.10}_{+0.09}$&0.30$^{+0.01}_{-0.01}$&
2.51$^{+0.02}_{-0.02}$&0.01&0.57$^{+0.00}_{-0.01}$&0.96\\
\hline
\multicolumn{9}{|c|}{PSR J0751+1807, $\nu$=287.4Hz, M$_{G}$=1.64$^{+0.15}_{-0.15}$}\\
\hline
6.66&0.76$^{+0.20}_{-0.12}$&1.81$^{+0.19}_{-0.18}$&13.76$^{-0.25}_{+0.08}$&0.25$^{+0.04}_{-0.03}$&
2.26$^{+0.20}_{-0.26}$&0.01&0.47$^{+0.04}_{-0.05}$&0.96\\
\hline
\multicolumn{9}{|c|}{PSR J2234+0611, $\nu$=279.3Hz, M$_{G}$=1.393$^{+0.013}_{-0.013}$} \\
\hline
5.36&0.60$^{+0.00}_{-0.00}$&1.51$^{+0.02}_{-0.02}$&13.85$^{-0.00}_{+0.00}$&0.20$^{+0.00}_{-0.00}$&1.82$^{+0.02}_{-0.02}$&0.01&0.36$^{+0.00}_{-0.00}$&0.95\\
\hline
\multicolumn{9}{|c|}{PSR J0437-4715, $\nu$=173.6Hz, M$_{G}$=1.44$^{+0.07}_{-0.07}$} \\
\hline
5.53&0.62$^{+0.04}_{-0.02}$&1.57$^{+0.08}_{-0.08}$&13.72$^{-0.01}_{+0.01}$&0.21$^{+0.01}_{-0.01}$&1.89$^{+0.13}_{-0.12}$&0.00&0.23$^{+0.02}_{-0.02}$&0.97\\
\hline
\end{tabular}
\end{table*}
\begin{table*}
\centering
\caption{\label{tab:tab06}The theoretically computed radii of 
catalogued in Table \ref{tab:tab04} for non-spinning configurations,
and the corresponding percentage bias in the allowed EOS  of hybrid stars
for model NHymQ4, if this model of  EOS
 is constrained using an accurate $\pm$5$\%$ for equatorial radius measurement. } 
\begin{tabular}{|c|c|c|c|c|c|}
\hline
No.&M$_{G}$(M$_{\odot}$)&$\nu$(Hz)&R(km)&R$_{eq}$&Bias($\%$)\\
\hline
1&1.667&465.1&13.47&14.07&42.6\\
\hline
2&1.41&416.7&13.57&14.12&38.4\\
\hline
3&1.4378&366.0&13.57&13.99&29.5\\
\hline
4&1.540&339.0&13.56&13.90&24.2\\
\hline
5&1.928&317.5&12.59&12.98&29.4\\
\hline
6&1.832&315.5&13.11&13.43&23.4\\
\hline
7&1.64&287.4&13.50&13.76&18.6\\
\hline
8&1.393&279.3&13.57&13.85&19.7\\
\hline
9&1.44&173.6&13.57&13.72&10.4\\
\hline
\end{tabular}
\end{table*}
\begin{table}
\caption{\label{tab:tab07} The binary neutron star masses (M$_{1}$(M$_{0}$),
$M_{2}$(M$_{0}$)) and corresponding radii (R$_{1}$(km), R$_{2}$(km)), 
dimensionless tidal deformabilities ($\Lambda_{1}$,$\Lambda_{2}$),
weighted dimensionless tidal deformability
$\tilde{\Lambda}$,tidal correction $\Delta\tilde{\Lambda}$ and chirp 
radius R$_{c}$(km) for EOS NHymQ4.}
\begin{tabular}{|c|c|c|c|c|c|c|c|c|c|c|c|}
\hline
M$_1$&M$_2$&R$_{1}$&R$_{2}$&$\Lambda_1$&$\Lambda_2$&$\tilde{\Lambda}$
&$\Delta\tilde{\Lambda}$&R$_{c}$\\
\hline
1.36&1.36&13.61&13.24&745.81&732.13&731&-4.65&8.87\\
1.38&1.35&13.24&13.24&689.22&791.99&729&13.38&8.86\\
1.40&1.33&13.13&13.24&624.36&865.55&732&33.79&8.87\\
1.42&1.31&13.24&13.24&576.46&937.39&728&49.59&8.86\\
1.44&1.29&13.24&13.24&526.85&1010.95&726&66.11&8.86\\
1.46&1.27&13.13&13.24&485.80&1093.60&729&81.93&8.86\\
1.48&1.26&13.24&13.34&446.46&1182.86&728 &98.46&8.86\\
1.50&1.24&13.34&13.24&408.83&1270.10&722&113.66&8.85\\
1.52&1.22&13.24&13.34&374.62&1354.77&715&127.10&8.83\\
1.54&1.21&13.24&13.34&345.54&1462.71&724&142.61&8.85\\
1.56&1.19&13.13&13.34&316.46&1570.30&717&157.44&8.83\\
1.58&1.18&13.13&13.34&289.09&1677.04&718&171.04&8.83\\
1.60&1.17&13.13&13.34&263.43&1777.80&704&182.64&8.80\\
1.61&1.16&13.43&13.57&245.63&1828.60&689&189.95&8.77\\
\hline
\end{tabular}
\end{table}

\pagebreak
\begin{figure}
\includegraphics[trim=0 0 0 0,clip,scale=0.4]{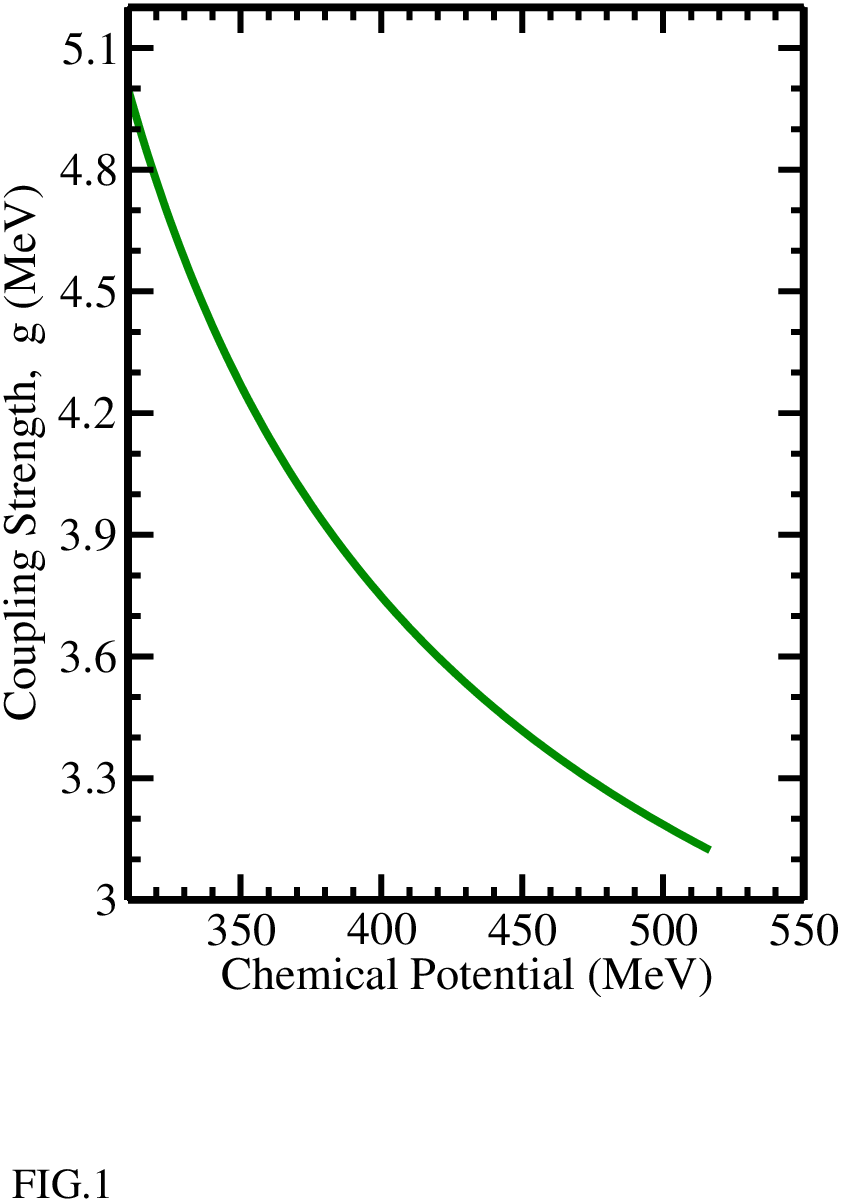}
\caption{\label{fig01} Variation in Coupling strength parameter with
chemical potential for Quarks Phase.}
 \end{figure}
 \begin{figure}
\includegraphics[trim=0 0 0 0,clip,scale=0.4]{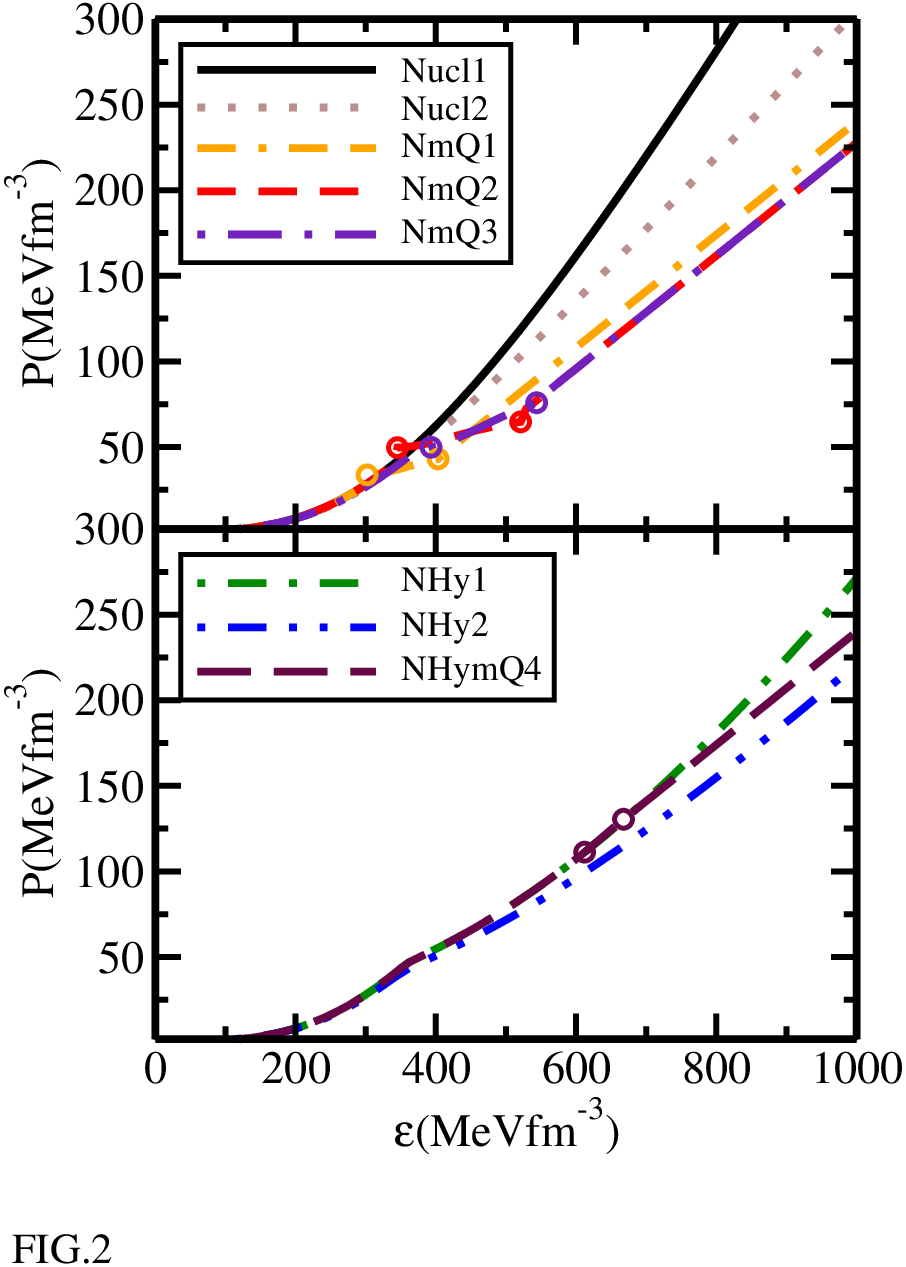}
\caption{\label{fig02}
The variation of pressure (MeVfm$^{-3}$) as a function of energy density
${\cal_{E}}$ (MeVfm$^{-3}$) for various selected EOSs.} 
\end{figure}
\begin{figure}
\includegraphics[scale=0.5]{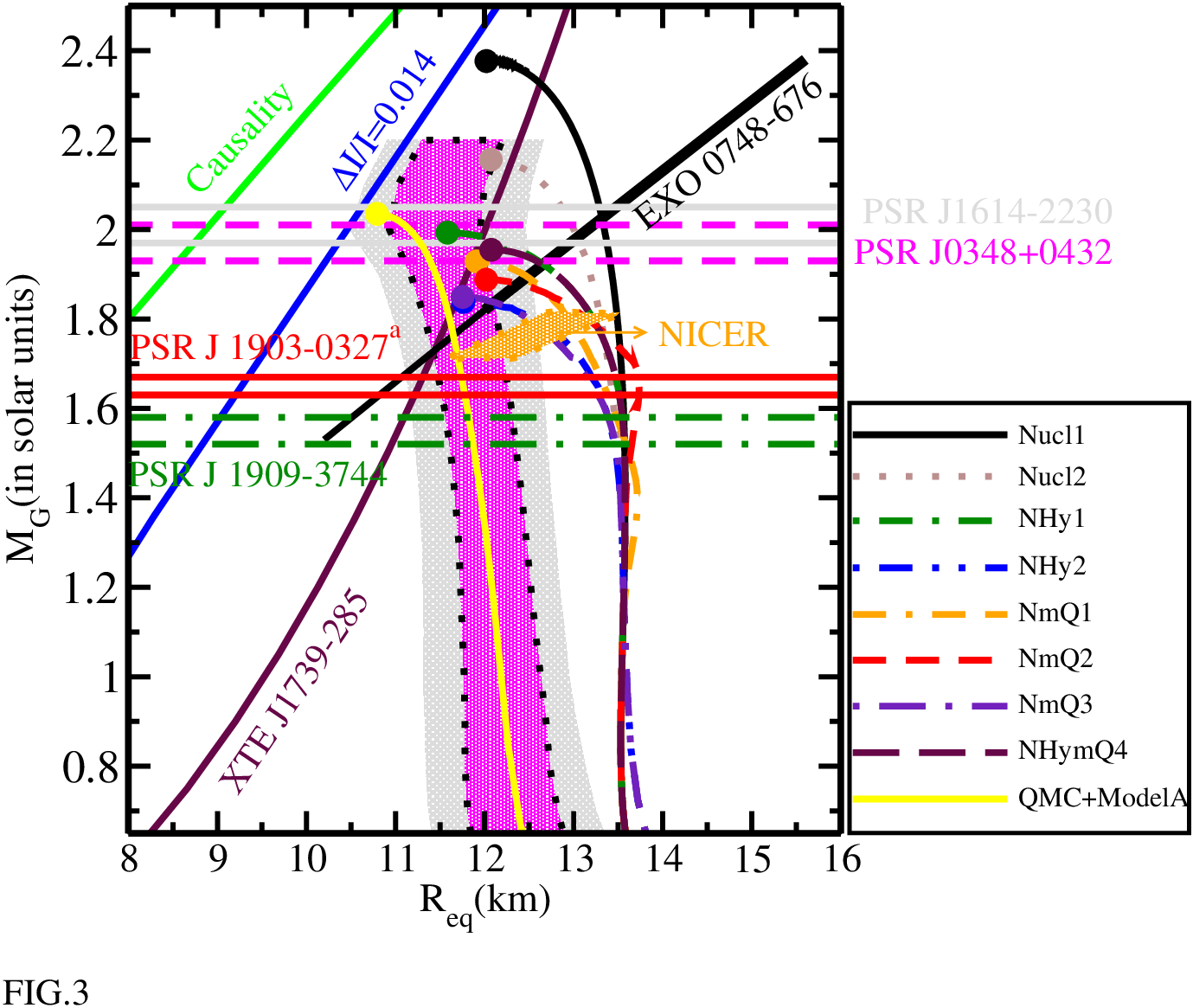}
\caption{\label{fig03} Relationship between gravitational mass and 
 radius of non - rotating compact star for various EOSs. The region excluded
 by causality light green solid line and rotation 
 constraints of neutron star XTE J1739-285 solid maroon line are given. 
 The mass and radius 
 limit estimated from Vela pulsar glitches $\Delta I/I$=0.014 is shown as
 blue solid line. The mass limits of pulsars
PSR J1614-2230 and PSR J0348+0432 are plotted for comparison. 
The limits on compact
star mass and radius from Ozel's analysis of EXO 0748-676 with 1$\sigma$ 
(dark solid black line) and 2$\sigma$ (extended black
line) error bars are also shown. The mass radius relationship obtained in
Ref.\cite{nattila2016} from extracted data of EOS by using QMC+Model A. 
The orange region bounded by the dotted maroon lines is representing the
mass - radius relationship extracted for the proposed pulsar PSRJ0437-4715 
for 3$\sigma$ confidence level in the NICER program \cite{Gendreau2012}.}
\end{figure}
\begin{figure}
\includegraphics[trim=0 0 0 0,clip,scale=0.35]{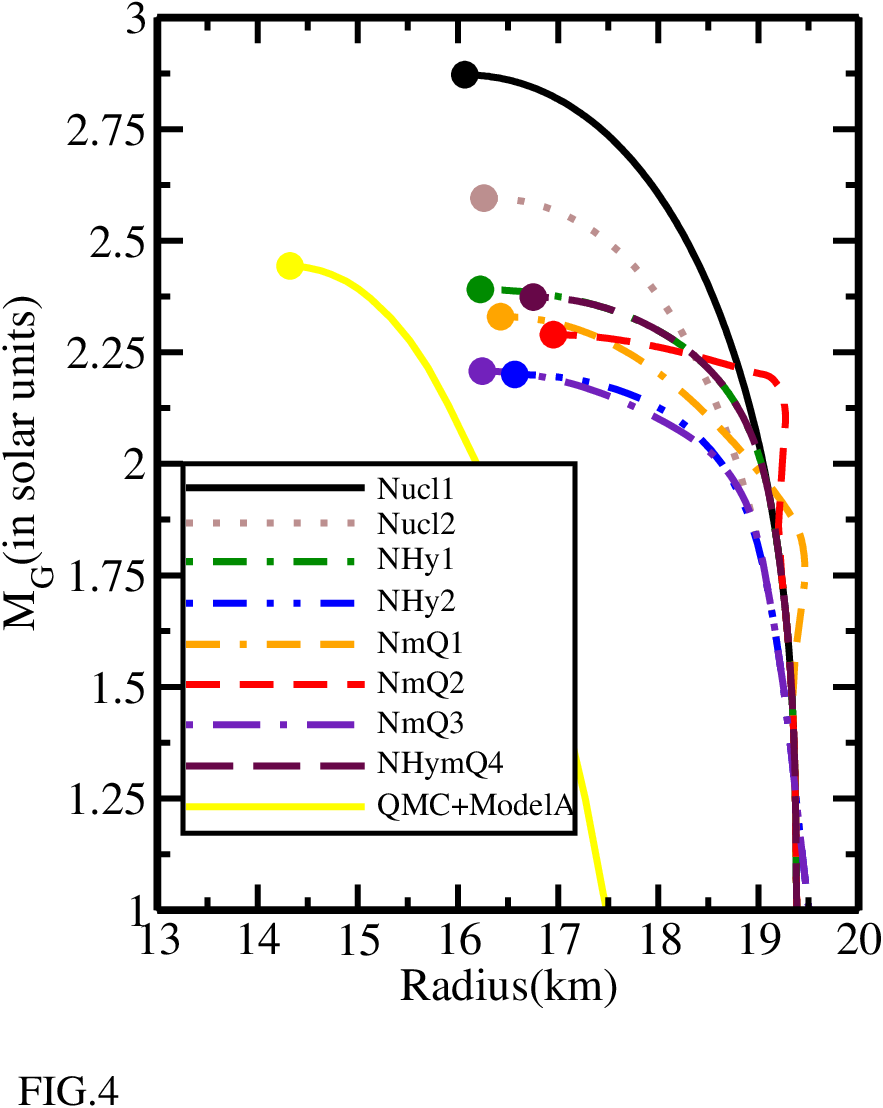}
\caption{\label{fig04} The variation of gravitational mass as a function
of equatorial radius with Keplerian configurations for various EOSs.}
\end{figure}
\begin{figure}
\includegraphics[trim=0 0 0 0,clip,scale=0.35]{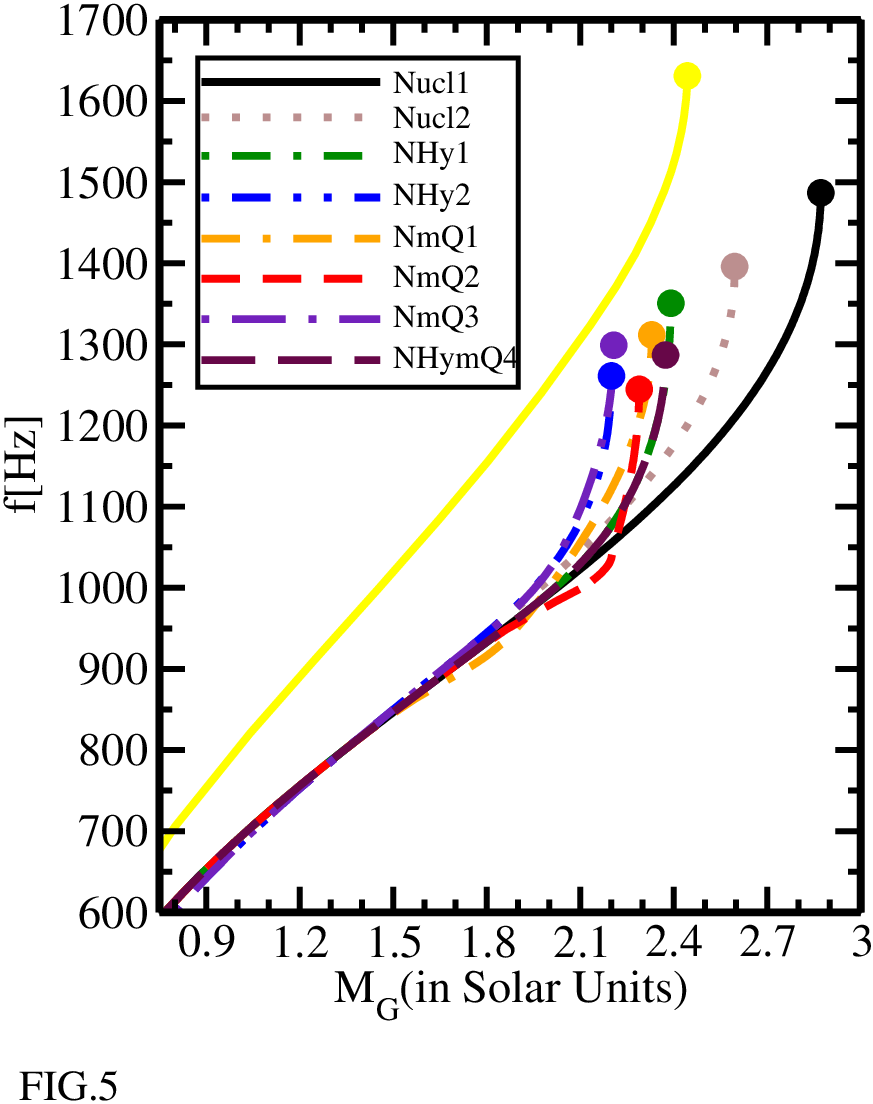}
\caption{\label{fig05} Keplerian frequency as a function of the gravitational mass.}
\end{figure}
\begin{figure}
\includegraphics[trim=0 0 0 0,clip,scale=0.35]{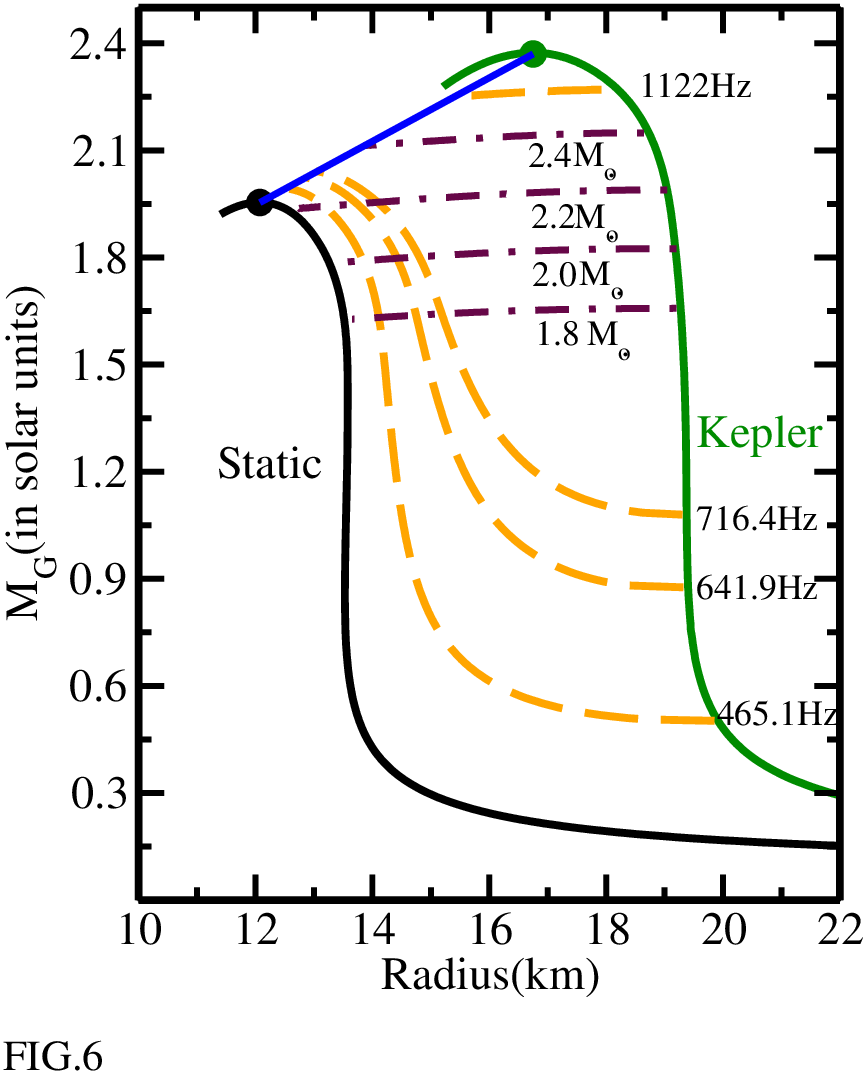}
\caption{\label{fig06} Gravitational mass (M$_{G}$)as a function of
the equatorial radius (R$_{eq}$) for rotating  hybrid star. Black solid
line in the figure presents static and dark solid green line presents 
the kepler sequences for NHymQ4 equation of state. The orange dashed 
lines are at constant frequencies 1122Hz, 716.4Hz, 641.9Hz and 465.1Hz
sequences.The horizontal dashed dotted line  represents the constant
baryon mass sequences at M$_{B}$=2.4M$_{\odot}$, 2.2M$_{\odot}$, 
2.0M$_{\odot}$, 1.8M$_{\odot}$.}
 \end{figure}
\begin{figure}
\includegraphics[trim=0 0 0 0,clip,scale=0.4]{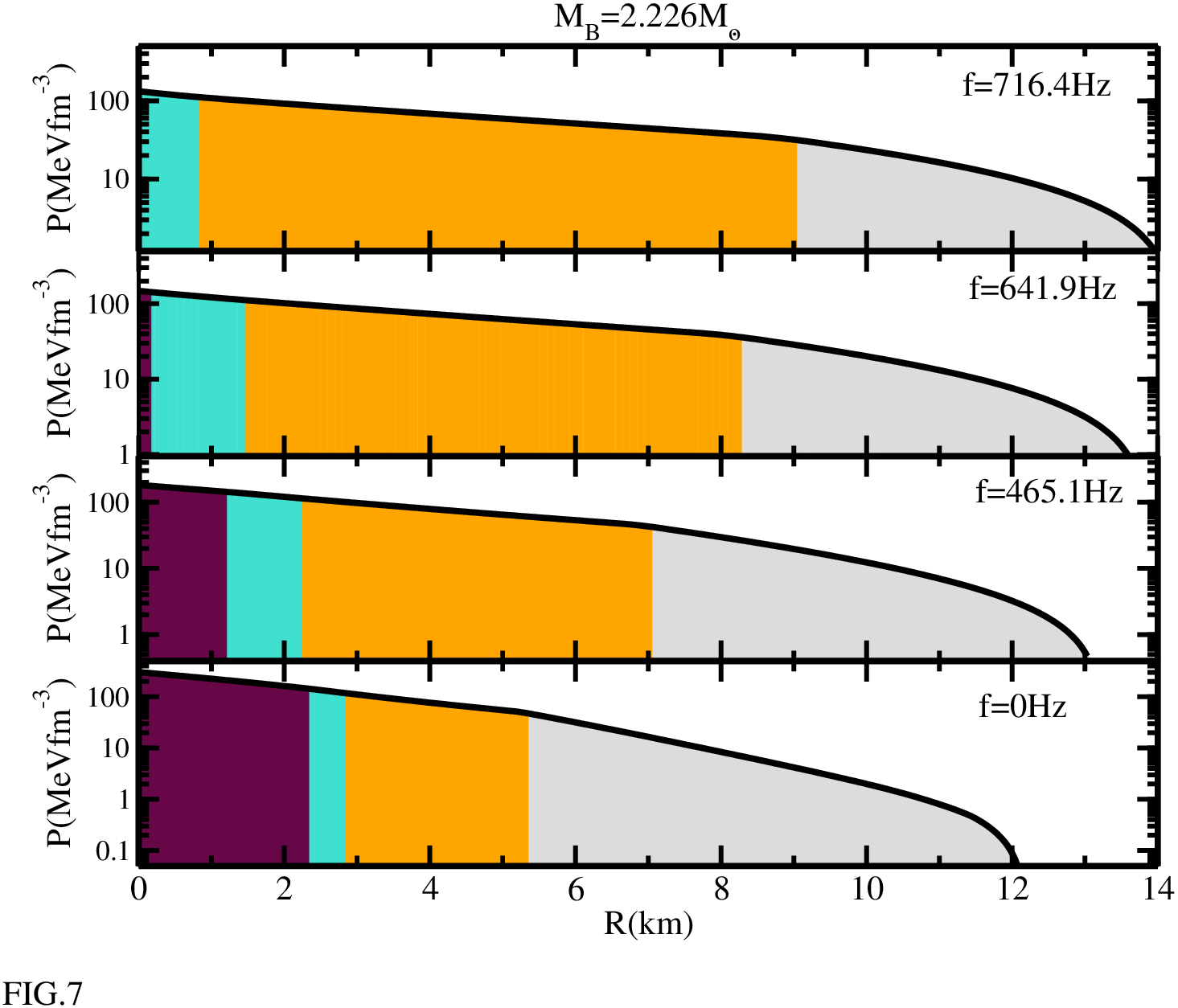}
\caption{\label{fig07}Equatorial profile of pressure as a function of
radial distribution for the non-rotating and spin-down sequences of constant 
baryonic mass M$_{B}$=2.226M$_{\odot}$ for hybrid star. The chosen spin-down 
frequencies(f) 716.4Hz, 641.9Hz and 465.1Hz are corresponding to observational
limits of the pulsars PSR J1748-2446ad\cite{Hessels2006}, PSR B1937+21\cite{Hessels2006}
and PSR J1903+0327\cite{Champion2008,Freire2011} respectively. 
The grey shaded region represented nucleonic phase only, the orange 
region represented Hadronic Phase, the turquoise region represented
the mixed phase of Hadrons, Hyperons and Quarks. The dark maroon
region represented pure quarks in CFL phase.}
 \end{figure}
 \begin{figure}
\includegraphics[trim=0 0 0 0,clip,scale=0.4]{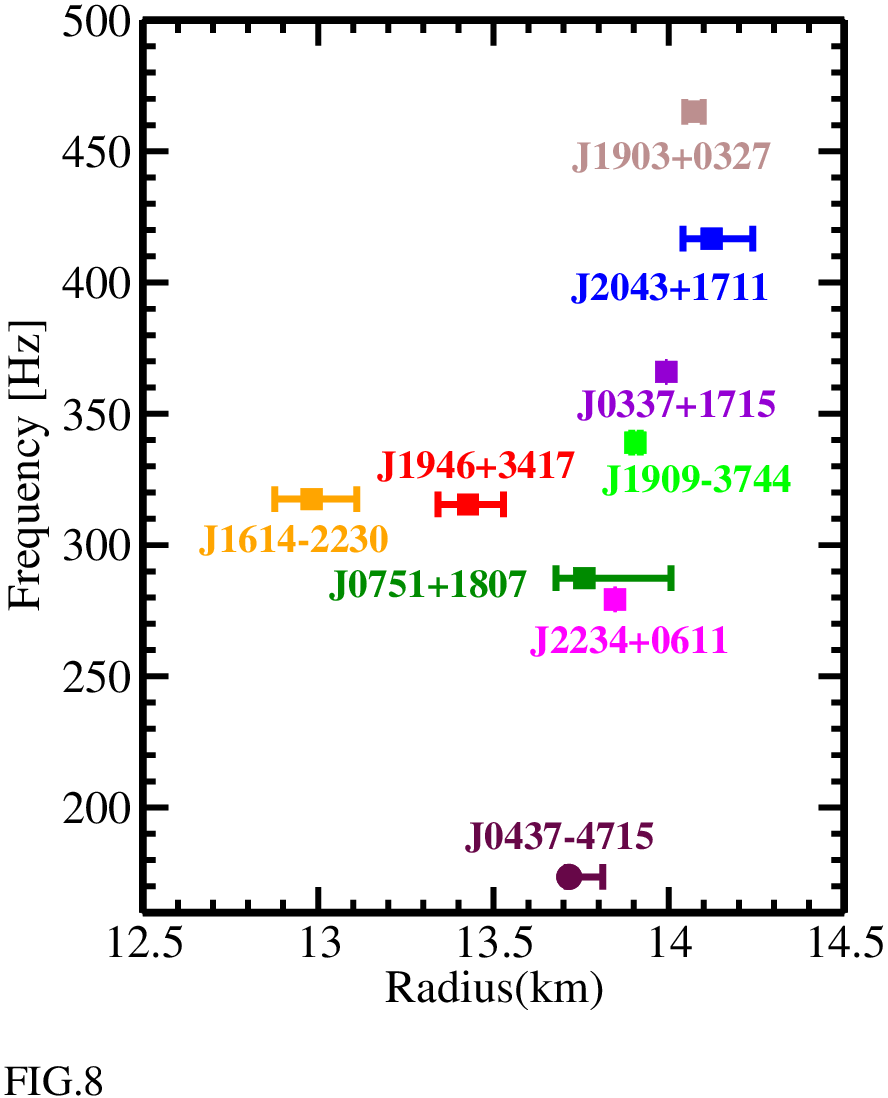}
\caption{\label{fig08} The theoretical limits of the radii of various well
known millisecond pulsars is presented as a function of their 
rotational frequencies. The theoretical limits of the radii and the 
gravitational mass of these pulsars have been computed by using NHymQ4
equation of state.}
\end{figure}
\begin{figure}
\includegraphics[trim=0 0 0 0,clip,scale=0.4]{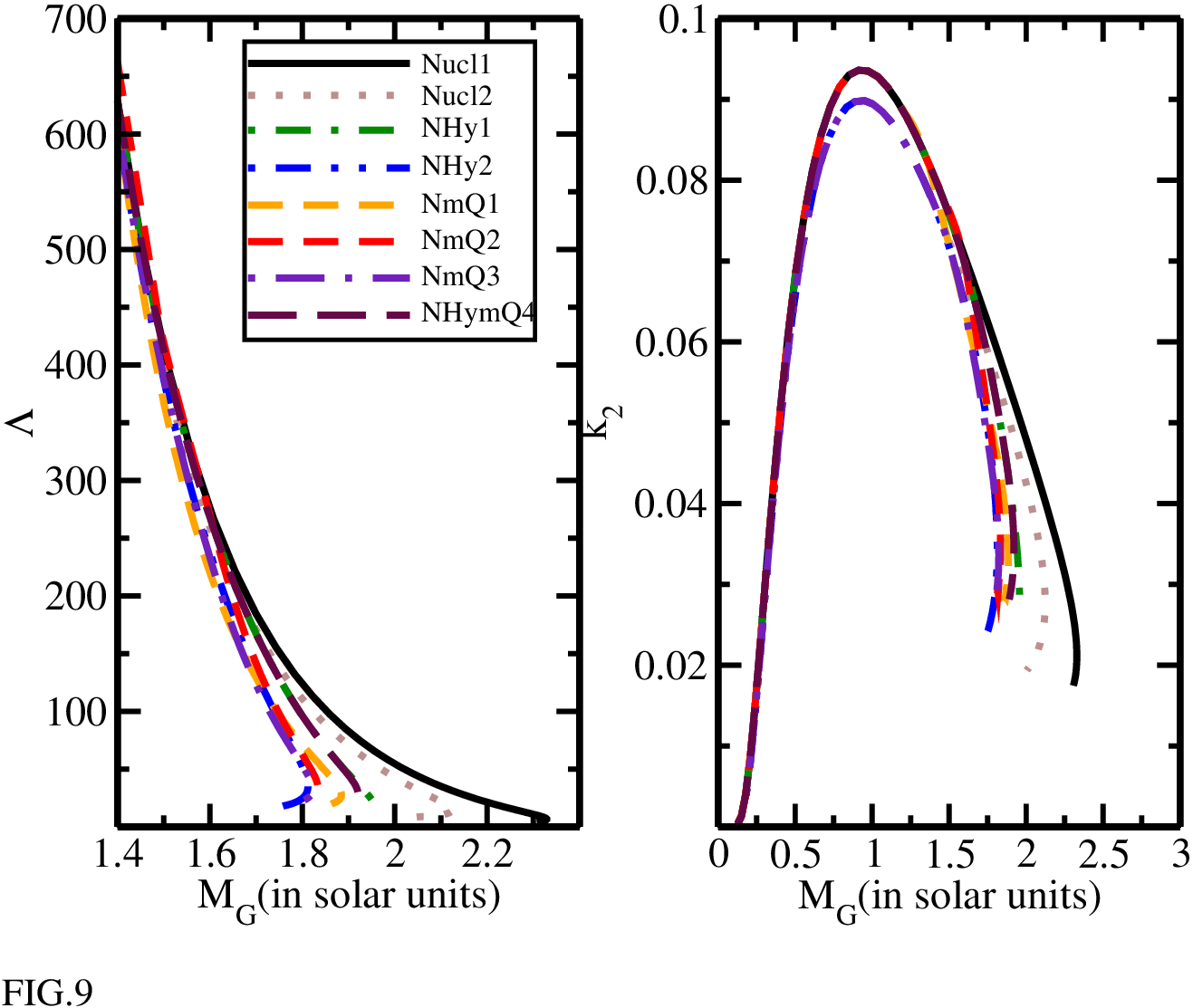}
\caption{\label{fig09} (Color online) (left panel) The tidal deformability ($\Lambda$)
and (right panel) the dimensionless Love number (k$_2$ ) with respect
to gravitational mass for different EOSs.}
\end{figure}
\begin{figure}
\includegraphics[trim=0 0 0 0,clip,scale=0.4]{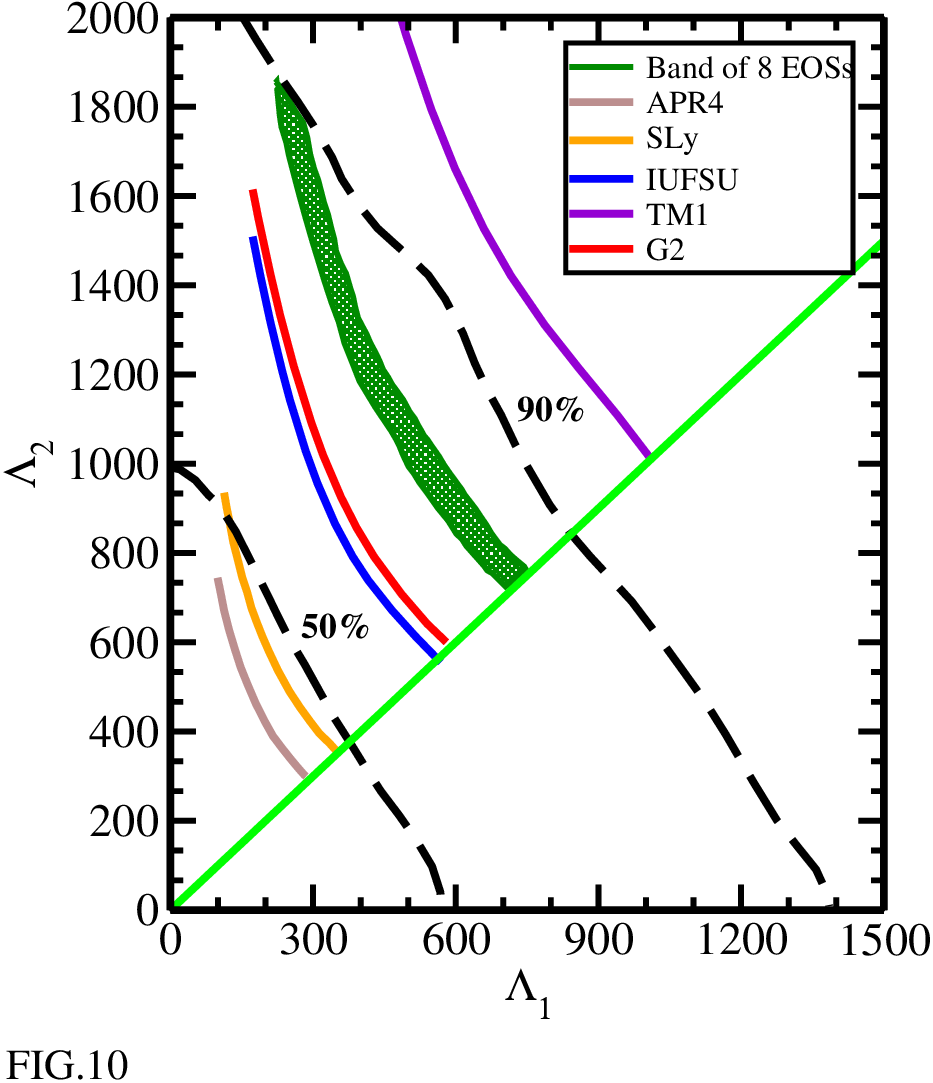}
\caption{\label{fig10} (Color online) Tidal deformabilities $\Lambda_1$ and
$\Lambda_2$ associated with the high-mass M$_1$ and low-mass M$_2$ components 
of the binary predicted by a set of eight EOSs. The 50$\%$ and 90$\%$ confidence
limits for this event are also indicated.}
\end{figure}

\end{document}